\documentclass{ar-1col}
\usepackage[numbers]{natbib}
\usepackage{url}
\usepackage{multicol}
\usepackage{cancel}
\jname{Xxxx. Xxx. Xxx. Xxx.}
\jvol{AA}
\jyear{YYYY}
\doi{10.1146/annurev-nucl-102419-055056}

\begin{document}

\markboth{G. Lanfranchi, M.~Pospelov, P. Schuster  }{The Search for Feebly-Interacting Particles}

\title{The Search for Feebly-Interacting Particles}

\author{Gaia Lanfranchi$^1$,
Maxim Pospelov$^{2,3}$ and
Philip Schuster$^4$
\affil{$^1$Laboratori Nazionali di Frascati dell'INFN, Frascati (Rome), Italy, 00044; email: Gaia.Lanfranchi@lnf.infn.it}
\affil{$^2$ School of Physics and Astronomy, University of Minnesota, Minneapolis, MN 55455, USA}
\affil{$^3$ William I. Fine Theoretical Physics Institute, School of Physics and Astronomy, University of Minnesota, Minneapolis, MN 55455, USA; email: mpospelov@perimeterinstitute.ca}
\affil{$^4$SLAC  National  Accelerator  Laboratory,  2575  Sand  Hill  Road,  Menlo  Park,  CA  94025,  USA; email: schuster@slac.stanford.edu}}

\begin{abstract}
At the dawn of a new decade, particle physics faces the challenge of explaining the mystery of dark matter, the origin of matter over anti-matter in the Universe, the apparent fine-tuning of the electro-weak scale, and many other aspects of fundamental physics. Perhaps the most striking frontier to emerge in the search for answers involves new physics at mass scales comparable to familiar matter, below the GeV-scale, but with very feeble interaction strength. New theoretical ideas to address dark matter and other fundamental questions predict such feebly interacting particles (FIPs) at these scales, and indeed, existing data may even provide hints of this possibility. Emboldened by the lessons of the LHC, a vibrant experimental program to discover such physics is under way, guided by a systematic theoretical approach firmly grounded on the underlying principles of the Standard Model. We give an overview of these efforts, their motivations, and the decadal goals that animate the community involved in the search for FIPs, with special focus on accelerator-based experiments. 
\end{abstract}

\maketitle

\tableofcontents

\section{INTRODUCTION}
\label{sec:introduction2}

{\it Feebly-interacting particles} (FIPs) have a notable history in fundamental physics. Indeed, weak-interaction phenomena provided physics with an early example of a FIP in the form of a neutrino, and more recently dark matter suggests the existence of additional types of FIPs. Aside from their empirical role, FIPs have also been common elements of our most popular theoretical ideas to extend the Standard Model, such as supersymmetry, and various low-energy limits of string theory. Even as particle physicists come to terms with null-results from the LHC, which constrains new physics (with sizable SM coupling) up to multi-TeV mass scales, many remaining theoretical ideas continue to predict FIPs in compelling ways. For example, FIPs provide excellent thermal dark matter candidates that extend the WIMP paradigm. The origin of neutrino masses and the matter/anti-matter asymmetry in the Universe are easily explained in extensions of the SM with FIPs. The simplest theories to explain the origin of $CP$ symmetry in strong interactions predict FIPs, and ideas to address the electro-weak hierarchy problem, and the origin of cosmological inflation, also predict FIPs. Given their prevalence in theoretical approaches to extending the SM, and empirical evidence for their existence, interest in FIPs has grown steadily over time. Such interest has only accelerated in recent years, spurred on by anomalies in astrophysical data, precision SM measurements, and by the profound lessons of the LHC. 

\vskip 2mm
Until a decade ago, most searches for FIPs were focused on electro-weak mass scales, in large part because they were expected as part of a broader range of new physics such as super-partners or extra dimensions. However, the bias towards these scales has been eroded steadily, both by LHC null results, and by the growing recognition that FIPs can address SM puzzles over a wider mass range that extends well below the GeV scale. In fact, the degree to which FIPs were poorly constrained below the GeV-scale caught physicists by surprise in many ways. Whereas below the MeV-scale, stellar and cosmological data constrains FIP couplings to be much smaller than that of weak-forces, in the MeV-GeV range, certain FIPs can interact {\it more strongly} than weak-forces and still be compatible with existing data! This fact is all the more striking given that stable SM matter resides in this MeV-GeV range, and given how important these scales are to BBN, the QCD phase transition, and other aspects of the early Universe. These combined realizations spawned the development of a new generation of experiments over the past decade, which has now grown to include fixed-target and collider accelerators of all varieties, direct and indirect detection experiments, and precision measurements of SM parameters. 

\vskip 1mm
This review aims to summarize the current state of FIP physics (often referred to as ``dark sector'' physics as well), focused on the relatively unexplored MeV-GeV mass range, with special attention given to accelerator-based experiments. Within this scope, our goal is to introduce particle physicists and astrophysicists to the essential motivations for FIPs, the most influential experimental techniques, and to explain the compelling science goals and milestones that energize the field as a whole. We will also highlight the interconnections and synergy among different experimental techniques including accelerators, DM direct detection, as well as astrophysics and cosmology. 

\vskip 1mm
As described in Section~\ref{sec:motivation}, much of the phenomenology relevant to FIP searches can be understood from basic principles of quantum field theory and symmetry, which powerfully constrains the way new (SM neutral) physics can interact with SM matter. Historically, this led to the birth of the ``portal formalism'' and the use of simple models to explore FIP-phenomenology, with the attitude that more complex models aimed at addressing broader SM puzzles would inevitably be covered by this approach. This expectation has been borne out in recent years, as more detailed models to address dark matter, the hierarchy problem, neutrino masses, and other puzzles have provided specific implementations of various simple models, and provide clues to help prioritize regions of parameter space deserving of special experimental attention. These studies have also brought into sharp focus a number of important scientific milestones that we will comment on throughout the review. Most of these milestones have focused on new physics mass scales at or {\bf below} the weak-scale, in the same mass range as familiar matter, and in the MeV-GeV range in particular.  

\vskip 1mm
As described in Sections~\ref{sec:experiments}-\ref{sec:exp_sensitivity}, the experimental program that has emerged to pursue FIP science is diverse, and the interest in this field is reflected in many recent community planning documents, such as the Cosmic~\cite{Feng:2014uja} and Intensity ~\cite{Hewett:2014qja} Frontier Reports of the 2013 Snowmass process, the Dark Sector community report~\cite{Alexander:2016aln}, the Cosmic Vision report~\cite{Battaglieri:2017aum}, the LHC Long-Lived Particle community white paper~\cite{Alimena:2019zri}, the Physics Beyond Colliders BSM report~\cite{Beacham:2019nyx}, the White Paper on new opportunities for next generation neutrino experiments~\cite{Arguelles:2019xgp}, and the Briefing Book of the European Strategy for Particle Physics~\cite{Strategy:2019vxc}. Searches for FIPs at accelerator-based experiments in particular have triggered tremendous activity, and are currently being performed at almost all the existing collider experiments and facilities, including ATLAS, Belle II, CMS, LHCb at colliders and HPS, miniBooNE, NA62, NA64, and SuperKamiokande at extracted beam lines. Moreover, a multitude of new initiatives to cover important parameter space have emerged at CERN (CODEX-b~\cite{Aielli:2019ivi}, FASER~\cite{Ariga:2018pin, Ariga:2018uku}, MATHUSLA~\cite{Alpigiani:2018fgd,Alpigiani:2020tva}, NA62 in dump mode~\cite{NA62:dump}, SHiP~\cite{Anelli:2015pba}), at SLAC (LDMX~\cite{Akesson:2018vlm}), at FNAL (DarkQuest~\cite{Batell:2020vqn}, DUNE Near Detector~\cite{Berryman:2019dme}, M3~\cite{Kahn:2018cqs}), at MESA (DarkMesa ~\cite{Christmann:2020qav}), at LNF (PADME~\cite{Raggi:2015gza}), at Jefferson Lab (BDX), and other laboratories. On a longer time scale, future high-energy electron and proton colliders will be able to explore higher mass ranges in a fully complementary way with respect to low energy searches. 

\vskip 1mm
In order to fully explore the types of sub-GeV FIPs (or dark sectors) motivated by dark matter, the strong CP problem, and other mysteries of the Standard Model, a healthy diversity of small to medium scale experiments operating at several facilities and employing a range of techniques will be required. Deriving meaningful scientific conclusions from these varied studies will require extensive collaboration and cross-fertilisation across different communities, some of which has already started in recent community planning initiatives. This is particularly true for sub-GeV DM searches -- these efforts range from searches in underground detectors via scattering of DM off of nuclei or atoms (direct detection), to observation of cosmic rays, gamma rays and neutrinos produced by annihilation of DM in the cosmos (indirect detection), to searches for missing energy/momentum/mass signals at accelerator-based experiments. 

\vskip 1mm
To help guide experimental efforts under way, the community should strive to use a common theoretical framework for FIPs and dark sectors that includes a representative set of minimal models built on clear theoretical principles of effective field theories. Adopting such a systematic approach will allow the community to combine/compare different experimental results. This is necessary for identifying promising and still uncovered regions of parameter space to guide future proposals/efforts, and also crucial for testing possible explanations in the case of positive signals.  

\vskip 1mm
In Section~\ref{sec:conclusions}, we comment in some detail on recent scientific achievements resulting from FIP searches over the past 10 years, and highlight next steps and milestones that the field is currently focused on. Some of these milestones are also called out in recent community prioritization efforts, such as the DOE BRN report~\cite{BRNReport}, and the Briefing Book of the European Strategy for Particle Physics~\cite{Strategy:2019vxc}. 

\section{MOTIVATIONS AND THEORETICAL FRAMEWORK}
\label{sec:motivation}

FIPs are any new (massive or massless) particle coupled to SM particles via extremely small couplings. The small strength of these couplings can naturally be due to either to the presence of an approximate symmetry only slightly broken and/or to the presence of a large mass hierarchy between particles. FIPs are neutral under the SM gauge interactions, although small weak neutral charge is a possibility. FIPs are fully complementary to new physics with sizeable couplings at the TeV scale explored (directly or indirectly) by the LHC experiments. Minimal models based on effective field theory have historically provided much of the guidance for exploring FIPs below the GeV-scale, and this will be explained in Section~\ref{ssec:portal_formalism}. But first, we summarize below the motivations for FIPs in the context of more general motivations for new physics beyond the Standard Model. 

\subsection{Motivations}
There is a multitude of models of FIPs that include additional new physics at higher mass scales, mostly driven by attempts to solve the Higgs hierarchy problem such as MSSM with R-Parity violation, Split-SUSY, Neutral Naturalness (see {\em e.g.} ~\cite{Barger:1989rk,Wells:2004di,ArkaniHamed:2004fb,Chacko:2005pe, Burdman:2006tz} for a representative set of ideas). While some of these ideas have certain merits, no undisputable top-down approach has emerged, and therefore in exploring the potential of FIPs as solutions to open problems in particle physics and cosmology we prefer to adopt a phenomenological viewpoint, reflected in our summary below. 

\vskip 2mm
\noindent {\bf The electro-weak hierarchy and strong CP puzzles.}\\
The incredible success of the SM in describing the vast majority of observable phenomena in Nature comes hand-in-hand with the SM appearing fine-tuned in striking ways. One of the most puzzling aspects of the SM is the vast hierarchy between the mass scale, governing the strength of the gravitational force, $M_{Planck}$ and the electroweak scale $m_H$: $m_H/M_{Planck}\propto 10^{-17}$. The Higgs particle has so far not shown any experimental signs of compositeness, suggesting that the associated quantum field is susceptible to quantum corrections that would drive its mass towards the highest known scale of new physics, which is presumably near $M_{Planck}$. The vast separation between the observed Higgs mass and $M_{Planck}$ therefore appears rather unnatural from a theoretical point of view. 

\vskip 1mm
A number of theories try to address this issue by constructing specific mechanisms for cancellations of large quantum corrections to the Higgs mass, such as {\em e.g.} super-symmetry. Nevertheless, it is possible that some other selection mechanisms are at play that explore different alternatives. These include a much lower cutoff for the gravitational interactions, such as in theories with large extra dimensions~\cite{ArkaniHamed:1998rs}, leading to Kaluza-Klein copies of tensor and scalar gravitons - which in essence signifies the emergence of large numbers of extremely weakly coupled FIPs below the EW scale. Some other ideas posit neutral naturalness \cite{Chacko:2005pe, Burdman:2006tz}, which imply some type of discrete symmetry that implies the existence of light particles in the "approximately mirror" sector, with extremely small couplings to the SM. Finally, it is conceivable that the Higgs mass was driven to its current value by some type of adjustment mechanism that exploits light scalar fields whose evolution drives Higgs mass to today's value \cite{Graham:2015cka}. 

\vskip 1mm
In many of these scenarios, the mechanism responsible for resolving the electro-weak hierarchy problem implies that FIP states $S$ couple to the Higgs in a manner described by, 
\begin{equation}
(H^\dagger H)\times m_H^2 \longrightarrow (H^\dagger H)\times( m_H^2 + c_1S+c_2S^2+...),
\end{equation}
and illustrate how models of new physics can realize FIPs coupled via the {\em Higgs portal} 
\cite{Patt:2006fw,Silveira:1985rk}. 

\vskip 1mm
Another puzzling aspect of the SM is the extreme smallness of the parameter $\theta_{QCD}$ that appears in front of gluon pseudo-scalar density, $\theta_{QCD}G_{\mu\nu}^a \tilde G^a_{\mu\nu}$,  which manifests itself in a number of non-perturbative phenomena. Chief among these are effects linear in $\theta_{QCD}$ that break $CP$ symmetry, and induce large (compared to experimental limits) electric dipoles moments of the neutron and heavy atoms \cite{Crewther:1979pi}. A FIP-type solution to this problem was found many years ago \cite{Peccei:1977hh,Weinberg:1977ma,Wilczek:1977pj}. Promoting $\theta$ to a new dynamical field (perhaps a Goldstone remnant of some additional global Peccei-Quinn symmetry), we have:
\begin{equation}
    \theta_{QCD}G_{\mu\nu}^a \tilde G^a_{\mu\nu} 
   \longrightarrow
    \left(\theta_{QCD}+\frac{a}{f_a}\right) G_{\mu\nu}^a \tilde G^a_{\mu\nu}.
\end{equation}
Non perturbative effects generate the new mass term that has 
$m_q\Lambda_{QCD}^3\left(\theta_{QCD}+\frac{a}{f_a}\right)^2$ dependence and ensures 
that the minimum of the potential restores $CP$ invariance of strong interactions. While original models had put $f_a$ close to the EW scale, $f_a \sim v $, it was later realized that the range for it is much larger, creating a vast landscape for the QCD axion mass and coupling. Moreover, enlarging the number of similarly generated axions particles and allowing for new $m_a$-generating mechanisms generalizes the QCD axion to a family of {\em axion-like particles}, or ALPs. 

\vskip 1mm
Finally, the gauge structure of the SM, the celebrated $SU(3)\times SU(2) \times U(1)$ group product, as well as the representations of SM matter fields are very suggestive of a unified gauge structure that in turn can have more low-energy remnants than the SM gauge group. Specifically, one may expect that 
\begin{eqnarray}
    (SU(3)\times SU(2) \times U(1))_{\rm SM}\longrightarrow 
    {\rm GUT~gauge~group} \nonumber \\
    \longrightarrow (SU(3)\times SU(2) \times U(1))_{\rm SM} \times U(1)_X\times...,
\end{eqnarray}
where  an additional (or several additional) $U(1)_X$ may be gauging additional accidental symmetries of the SM, such as $B-L$, or be entirely new "dark groups" with very small couplings to the SM fields. If the mass scale for the additional $U(1)_X$ is small, the new gauge bosons and additional matter fields are also a motivated case for FIPs. 

\vskip 2mm
\noindent {\bf Neutrino oscillations imply a new matter sector.}\\
Precise measurements of neutrino flavor oscillations point to the existence of neutrino masses, and a mismatch between weak and mass eigenstate bases. The non-zero neutrino mass dictates the existence of new states that participated in generating it. Among various neutrino mass generation mechanisms, the one that is based on a right-handed neutrino field $N$ is the most economical and most natural, both for Dirac (D) and Majorana (M) neutrinos:
\begin{eqnarray}
m_{\nu,\rm D}\bar \nu \nu  &&\longrightarrow y_\nu \bar N \nu H  + (h.c.)\\
m_{\nu,\rm M}\bar \nu \nu &&\longrightarrow (y_\nu)^2 (\nu H)^c \times \frac{1}{m_N} \times (\nu H) + (h.c.)
\end{eqnarray}
The Dirac case is a clear example of a FIP, with a new field $N$ sharing the same observable neutrino mass in the meV-to-eV range, and implying the size of the Yukawa coupling as small as $10^{-13}$. The Majorana case features a much heavier particle $N$ that we would refer to as Heavy Neutral Leptons (HNLs), and $m_N^{-1}$ in the mass generation mechanism (also known as "see-saw") is the propagator of the HNL, $\langle N\bar N \rangle$. The possible mass range for the $m_N$ is vast, and moreover, the see-saw scaling \cite{Minkowski:1977sc}
$m_\nu \propto y_\nu^2v^2m^{-1}_N$ does not necessarily have to hold due to the existence of multiple generations of HNLs and hidden symmetries among mass and Yukawa parameters. 

\vskip 1mm
It is very intriguing that the Majorana mass term for HNLs breaks the Lepton number by two units, and together with $B+L$ breaking provided by the non-perturbative electroweak effects at high temperatures, the HNLs offer an attractive path to a dynamical generation of matter anti-matter asymmetry \cite{Kuzmin:1985mm,Fukugita:1986hr}. This scenario, known as leptogenesis, was shown to be viable both with heavy states, but also with the HNLs below the EW scale \cite{Akhmedov:1998qx}, in which case they become perhaps the most motivated example of light fermionic FIPs.

\vskip 2mm
\noindent {\bf Cosmology and astrophysics requires new physics.}\\
While so far we have discussed subtle observational effects (neutrino flavor oscillations, (non)conservation of $CP$ invariance in strong interactions) or theoretical problems of the SM, it is data from cosmology and astrophysics that provide the most urgent evidence for new physics. The scientific revolution in cosmology lasting over the past 25 years has brought certainty, and sometimes extreme precision, to our knowledge of the history and composition of the Universe. The 100\% asymmetry between matter and antimatter, the necessity to generate the observable matter fluctuation spectrum, and above all the dominance of the dark sector energy density over "ordinary" matter, all point to new physics beyond the Standard Model. 

\vskip 1mm
The dark matter, known to be cold, and to a certain extent collisionless, had to be in existence prior to recombination. We know its abundance rather well, $\Omega_{\rm DM} \simeq 0.27$, which exceeds the abundance of atoms by a factor of 5.4. Other than that, its origin is a complete mystery, apart from the knowledge that it is {\em not} composed of any known SM particles. While "macroscopic" forms of DM are possible ({\em e.g.} primordial black holes), it is very likely that remnants of the hot Big Bang account for the DM abundance. One can point to several generic classes of theoretical ideas, each with its own merit, that provide a consistent history for DM's creation. As we will see, for each of these ideas, FIPs play an important role. 

\vskip 1mm
{\em Freeze-out DM, or WIMPs.}
Even very small couplings of DM to SM particles can lead to efficient thermalization in the early Universe. Subsequent expansion and cooling leads to the depletion of the DM energy density via {\em e.g.} 
${\rm DM +DM}\to {\rm SM} $ annihilation processes. In this case, the observed abundance of dark matter is set almost exclusively by the annihilation cross section, and is largely insensitive to unknown details of the early Universe, and to its mass. A thermally averaged $\langle \sigma v \rangle$  cross section,
$\langle \sigma v \rangle = (2-4) \times 10^{-26} \, {\rm cm}^3 {\rm s}^{-1} \simeq 1 \, {\rm pb} \cdot c$ leads to a relic freeze-out abundance of stable DM particles that matches the observed DM fraction of the total average energy density in the Universe, $\Omega_{\rm DM} \simeq 0.27$.

The canonical example of a minimal SM extension that realizes this scenario involves a heavy particle with mass between [0.1-1] TeV interacting through the weak force (Weakly interacting Massive Particle or WIMP). However a thermal freeze-out origin is naturally valid even if DM is at lower mass: DM with any mass in the MeV-100 TeV range can achieve the correct relic abundance by annihilating directly into SM matter. Light, {\em i.e.}  MeV-GeV range WIMPs would be overproduced in the early Universe, 
unless they are accompanied by an additional SM neutral mediators to enhance their annihilation rate~\cite{Boehm:2002yz,Boehm:2003hm,Pospelov:2007mp,Pospelov:2008zw,ArkaniHamed:2008qn, Pospelov:2008jd}. These “dark sector mediators" could be light, long-lived, vector or scalar FIPs that do not carry electromagnetic charge. This simple possibility that dark matter is a stable particle interacting via FIPs (sometimes referred to as dark sector dark matter) is one of the most predictive motivations underlying experimental searches for FIPs. 

\vskip 2mm
{\em Freeze-in DM.} What if the DM interaction with the SM is so weak that full thermalization in the early Universe never occurs, while the initial abundance of such particles is negligible?  The DM may be generated via ${\rm SM}\to {\rm DM} $ or ${\rm SM}\to {\rm DM+DM} $ processes. The number density of such particles remain subdominant to the leading SM species ({\em e.g.} photons) at all times. The requirement of being sufficiently cold typically restricts masses of such particles above $m_{\rm DM}\sim$keV, while the size of couplings preventing thermalization has to be exceedingly small (often $O(10^{-10})$ and smaller). This is a fairly flexible framework with many of the FIPs (and only FIPs!) being suitable for the role of the freeze-in dark matter, including the HNLs. 

\vskip 2mm
{\em Bosonic condensate DM.} As is well known, bosonic particles can have huge occupation numbers, vastly exceeding a naive thermal estimate $\propto T^3$ for their number densities. This is a non-thermal starting point for a whole class of DM models that include the QCD axion, which is among the most motivated examples in this category. However, in general, the mass scale for such DM candidates may be almost arbitrary, with the only real restriction being that their coupling to the SM must be small. Not surprisingly, this whole class of DM candidates relies on DM to be a FIP. 

\vskip 1mm
Thus we see that all three main classes of DM ideas feature FIPs, and for the rest of this review we concentrate on freeze-out DM, as the prospects for its discovery/exclusion in laboratory experiments are the greatest. We also would like to mention that the generation of primordial fluctuations via inflation (which remains to be the most compelling option), does not specify the mass scale for the main actor, the inflaton. This scalar field can be light, and much lighter not only compared to the Hubble expansion rate during inflation, but also compared to the electroweak scale. 

\vskip 1mm
To conclude this short cosmological discussion, we note that despite a seemingly endless number of possibilities for the solutions of cosmological problems ({\em e.g.} dark matter problem), there are also severe restrictions born out of observations and their comparisons to $\Lambda$CDM predictions. For example, the initial stages of structure formation, recombination, and cosmic microwave background (CMB) decoupling were occurring in a "quiet Universe" where the non-thermal injection of ionizing radiation was reduced to a minimum (below $\sim0.01-0.1$ eV per baryon). This puts severe pressure on light WIMP models, should they continue their residual annihilation with a ${\rm pbn}\times c$ rate at late times. Also well known is that the cosmological expansion during the big bang nucleosynthesis (BBN) and the CMB epochs appear to be normal, not showing any evidence for an additional thermalized degrees of freedom. Taken at face value, it disfavors light thermalized DM with a mass of a few MeV or less. These constraints help to narrow down the spectrum of viable FIP models to masses above the MeV scale. 

\vskip 1mm
A lively experimental activity aimed at searches of cosmologically motivated FIPs, including light DM and corresponding light mediators, 
is currently ongoing at almost all the accelerator-based experiments in the world. The current status of the searches and projections for future proposals will be described in detail in Section~\ref{sec:experiments} after having provided in the next Section (Section~\ref{ssec:portal_formalism}) a coherent theoretical framework and the explanation of the corresponding parameter space.

\subsection{Theoretical framework}
\label{sec:theoretical framework}

In this Section we provide an overview of the theoretical framework commonly used to describe the phenomenology of FIPs, starting with a summary of a general effective field theory formalism (i.e. portal formalism), and then a more detailed discussion of each of the major classes of FIP interactions with SM matter. 

\subsubsection{The Portal Formalism}
\label{ssec:portal_formalism}

To begin, consider new particles that do not carry electromagnetic, weak, or strong interaction charges. Such particles defined this way are often referred to as a {\it hidden sector} or {\it dark sector}. Let $O_{\rm SM}$  be an operator composed of SM fields, and $O_{\rm DS}$ an operator composed from the dark sector fields. The {\it ``portal framework"} (see {\em e.g.} Refs.~\cite{Patt:2006fw,Batell:2009di,Alekhin:2015byh, Beacham:2019nyx}) refers to a systematic way to describe all such interactions between a dark sector and the SM by building an interaction Lagrangian out of products of such operators, 
\begin{equation}
{\cal L}_{\rm total} = {\cal L}_{\rm SM } + {\cal L}_{\rm DS }
+ {\cal L}_{\rm portal};~~
{\cal L}_{\rm portal} = \sum  O_{\rm SM}\times O_{\rm DS}
\end{equation}
where the sum goes over a variety of possible operators and of different composition and dimension. Notice that only the product of the operators must be a Lorentz scalar, while $O_{\rm SM}$ and $O_{\rm DS}$ can transform as {\em e.g.} a spinor, or Lorentz tensor. However, given the assumption of the SM neutrality of the dark sector, both types of operators separately must be complete gauge singlets, and this places a powerful constraint on the types of interactions that are allowed.  

\vskip 1mm
As commonly defined, the minimal {\it "portals"} are the collection of lowest canonical-dimension operators that mix new dark-sector states with gauge invariant (but not necessarily Lorentz-invariant) combinations of SM fields. Following the general principles stated earlier, it turns out that the collection of such portals is rather simple, shown in Table 1. Finally, while 
${\cal L}_{\rm portal} $ has a rather minimal number of options, the dark sector Lagrangian itself can be far more complicated, and $A',S,N,a$ can connect, even at the renormalizable level to completely new particles and fields. 

\begin{table}
\caption{The portal formalism.}
\label{tab:portals}
\begin{center}
\begin{tabular}{rl}
Portal & Coupling \\
\hline
\smallskip
Vector: Dark Photon, $A'$ &  $-{\varepsilon \over 2 \cos\theta_W} F'_{\mu\nu} B^{\mu\nu}$ \\ \smallskip
Scalar: Dark Higgs, $S$  & $(\mu S + \lambda_{\rm HS} S^2) H^{\dagger} H$  \\ \smallskip Fermion: Heavy Neutral Lepton, $N$ & $y_N L H N$ 
\\ \smallskip Pseudo-scalar: Axion, $a$ &  ${a \over f_a} F_{\mu\nu}  \tilde{F}^{\mu\nu}$, ${a\over f_a} G_{i, \mu\nu}  \tilde{G}^{\mu\nu}_{i}$,
${\partial_{\mu} a \over f_a }\overline{\psi} \gamma^{\mu} \gamma^5 \psi$
\\
\end{tabular}
\end{center}
\end{table}

\vskip 1mm
In Table~\ref{tab:portals}, $F'_{\mu \nu}$ stands for the field strength of a dark gauge group, which couples to the hypercharge field, $B^{\mu\nu}$; $S$ is a new scalar singlet that couples to the Higgs doublet, $H$, with dimensionless and dimensional couplings, $\lambda_{\rm HS}$ and $\mu$; and $N$ is a new neutral fermion (or HNL) that couples to one of the left-handed doublets of the SM and the Higgs field with a Yukawa coupling $y_N$. These three cases capture most of UV-complete portal interactions which are, as such, unsuppressed by any possible large (or very large) new physics scale $\Lambda$. We will discuss possible enlargement of this set in Section~\ref{ssec:beyond_portals}. These three interactions, neutral under all the SM symmetries, keep the EW sector of the SM completely intact. While many new operators can be written at the non-renormalizable level, a particularly important example is provided by the axion (or axion-like) particle $a$ that couples to gauge and fermion fields at dimension-5. This is described by the pseudo-scalar portal which is suppressed by $1/f_a$, where $f_a$ is called the axion decay constant.

\vskip 1mm
SM symmetries have a number of important built-in properties such as minimal flavour violation, in which all flavor violation is associated only with the SM Yukawa matrices, minimal Higgs content, and separate perturbative conservation of lepton and baryon number. The first two portals in Table 1 fully preserve this minimal SM flavor structure, and therefore are not constrained by {\em e.g.} $K$, $B$ meson mixing and/or charged lepton conservation.
On the other hand, the HNL interactions do introduce new flavour dependence, via $y_N$ and/or the mass term for the HNLs themselves. This mass term - if Majorana - will also effectively break the lepton number by two units, making it a unique feature of this portal.   
Finally, the axion interaction with the SM fermions can source flavour violation both in the lepton and in the quark sectors. 
A more detailed description of the four portals is reported in the following.

\vskip 2mm
\subsubsection{Vector Portal}
\label{sssec:vector_portal}

A considerable amount of the theoretical and experimental attention of recent years has focused on the vector, or dark photon portal, with a distinct possibility that the new matter fields charged under the associated $U(1)_{A'}$ provide a viable thermal dark matter candidate. The minimal dark sector Lagrangian can be written as,
\begin{equation}
\label{A'}
    {\cal L} = -\frac14 F'_{\mu\nu }F'_{\mu\nu} -\frac{\epsilon}{2} F'_{\mu\nu }F'_{\mu\nu} +A_\mu' J^{\rm dark }_\mu+{\cal L}_{\rm mass}+...,
    \label{eq:L_vect}
\end{equation}
where the ``dark current" is given by the interaction of the $A'$ with either new fermions or new scalars: $J^{\rm dark }_\mu = g_{\rm D}\bar\chi \gamma_\mu \chi;~ ig_{\rm D}(\chi^* \partial_\mu \chi-\chi \partial_\mu \chi^*) $. The mass Lagrangian may include $\frac12m_{A'}^2(A'_\mu)^2$, $m_\chi\bar \chi \chi$ and $m_\chi^2\chi^* \chi$ terms, as well as interactions that could introduce spontaneous $U(1)_{A'}$ breaking ({\em i.e.} dark Higgs mechanism) and/or mass splitting of $\chi$ matter fields. 
The exchange by the mixed photon-dark photon propagator creates an interaction between the SM electromagnetic current and dark currents, 
\begin{equation}
\label{int}
 V_{\rm int}(q) =  J_\mu^{\rm EM} 
 \frac{\varepsilon}{q^2-m_{A'}^2}J_\mu^{\rm dark }
\end{equation}
which is suppressed by the small {\it kinetic mixing} parameter $\varepsilon$. We note that taking an $m_{A'}\to 0 $ limit corresponds to giving $\chi$ particles an EM "milli-charge", $Q_\chi = g_{\rm D} \times \varepsilon$. In the above, we have neglected the induced coupling of the dark current to the SM weak-current because it is additionally suppressed by $\left( m_{A'}/m_{Z} \right)^2$, and can therefore often be omitted at low energies, 

\vskip 1mm
The set of minimal models connected to the vector portal are theoretically appealing in several respects. The smallness of $\varepsilon$ can result from radiative effects where loops of matter charged under the SM gauge groups and $U(1)_{A'}$ will naturally generate a non-zero mixing angle $\varepsilon$, while $g_{\rm D}$ is not required to be small. Depending on the details of the loop-level UV physics, $\varepsilon $ could vary in a very wide range, {\em e.g.} $10^{-15}-10^{-1}$, though the naive one-loop estimate gives $\varepsilon \sim 10^{-4}-10^{-2}$, and GUT-inspired supersymmetric models tend to give $\varepsilon \sim 10^{-6}-10^{-3}$~\cite{ArkaniHamed:2008qp,Baumgart:2009tn}. 

\vskip 1mm
Perhaps the greatest appeal of this class of models is that they easily accommodate a minimally coupled WIMP-like dark matter candidate. Interaction (\ref{int}) allows for the annihilation of two dark particles to a pair of charged SM fermions, $\chi\bar\chi \to \psi_{\rm SM}\bar\psi_{\rm SM}$, or more generically to all combinations of the SM particles that can be created from the vacuum by the EM current. The CMB bounds on late energy injection \cite{Slatyer:2009yq} are easily evaded if the annihilation occurs in $p-$wave (such as for scalar or Majorana $\chi$), or if the mass splitting in the $\chi-\bar\chi$ system cuts off the late time annihilation \cite{Izaguirre:2013uxa}.

\vskip 1mm
The frequently analyzed parameter space for this model consists of \{$\alpha_D, \epsilon, m_{A’} ,m_{\chi}$ \} where $\alpha_D = g^2_{\rm D}/4 \pi$ is the dark coupling between DM and the vector mediator. Moreover, for achieving the correct dark matter abundance, a specific combination $y$ of these parameters is introduced \cite{Izaguirre:2013uxa},
\begin{equation}
    y = \alpha_D \varepsilon^2 \alpha \left( \frac{m_{\chi}}{ m_{A'}}\right)^4,
    \label{eq:y}
\end{equation}
that away from the $2m_\chi \simeq m_{A'}$ resonance more or less directly determines the annihilation rate, $\langle \sigma v \rangle \sim y\times m_{\chi}^{-2}$, and hence the final relic density. Tuning this rate to the one required to achieve the observed DM abundance defines a preferred range (or "thermal target") on the $\{m_\chi,y\}$ parameter space. 

\vskip 1mm
It is clear that Lagrangian (\ref{A'}) and the interaction (\ref{int}) provide a basis for experimental studies of FIPs coupled via the vector portal. Production of the $A'$ (or DM particle/anti-particle pairs $\chi\bar\chi$), with subsequent decay to SM or dark particles are among several distinct possibilities for exploring the vector portal in a wide variety of on-going and planned experiments. 

\subsubsection{Scalar Portal}
\label{sssec:scalar_portal}

The discovery of the Higgs boson $h$, prompts to investigate
the so called scalar or Higgs portal, that couples the dark sector to the Higgs boson via the bilinear $H^{\dagger}H$ operator of the SM. 
The minimal scalar portal model operates with one extra
singlet field $S$ and two types of dimensional and dimensionless couplings, $\mu$ and $\lambda_{\rm HS}$, respectively, 
\begin{equation}
 L_{\rm scalar} = L_{\rm SM} + L_{\rm DS} - (\mu S + \lambda_{\rm HS} S^2)H^{\dagger}H,    
\end{equation}
and can be easily generalized to multiple scalars. Intriguingly, the $\mu$-proportional portal belongs to the class of super-renormalizable operators, and is unique in that respect among all portal couplings that we consider. 

\vskip 1mm
The dark sector (DS) Lagrangian may include the interaction with dark matter $\chi$, 
$L_{\rm DS} = S \chi \overline{\chi} ..+....$. Most viable DM models in the sub-EW scale range imply $  m_{\chi} > m_S$,  
hence a {\it secluded annihilation} \cite{Pospelov:2007mp} $\chi \chi \to S S $ via a $t-$channel, $p-$wave transition. 

At low energy, the Higgs field can be substituted for $H = (v+h)/\sqrt{2}$, where $v$ = 246 GeV is the the EW vacuum expectation value, and $h$ is the field corresponding to the physical 125 GeV Higgs boson. The non-zero $\mu$ leads to the mixing of $h$ and $S$ states. In the limit of small mixing it can be written as

\[
 \theta = {\mu v \over m_h^2 - m_S^2 }.
\]
Hence the hidden scalar couples to SM fermions and vector bosons as a SM Higgs but with a strength reduced by a factor of $\sin \theta$, being $\theta$ the mixing angle between the two sectors. 

\vskip 1mm
The coupling constant $\lambda_{\rm HS}$ leads to the coupling of $h$ to a pair of $S$ particles, $\lambda_{\rm HS} S^2$. It can lead to pair-production of $S$ but cannot induce its decay. Hence in this case the scalar singlet does not mix with the Higgs, as in the case of an exact $Z_2$ symmetry. By itself, such scalar with the $Z_2$ symmetry was often considered as a candidate for DM \cite{Silveira:1985rk}. However, the experimentally observed Higgs particle with no detectable invisible decay channel, and new advances in direct detection together relegate the surviving mass range to a multi-TeV region. 

\vskip 1mm
An important property of the scalar portal, and the crucial difference with the vector portal is in the phenomenology of flavour-changing neutral currents (FCNC). While both portals preserve flavour at tree level, the $t-W$ loop-induced FCNC amplitudes for $b\to s,d$ and $s\to d$ transitions behave in a markedly different way. Due to conservation of the EM current, the dark photon penguin amplitudes are effectively suppressed (aside of the usual loop factors and CKM matrix elements) by
$G_Fm_{\rm meson}^2$. In the scalar portal case there is no current conservation, and an analogous combination scales as 
$G_Fm_{t}^2$, leading to an enormous enhancement factor. 
In practical terms, it means that the production of light FIPs via scalar portal will occur via $K$ and $B$ mesons. 
This makes experiments with copious flavored particle production, such as high-luminosity $e^+e^-$ $B$-factories, kaon and neutrino beam experiments, the LHC and hadronic beam dumps at high enough energy to be particularly suited for exploring the scalar portal in the low mass range.

\vskip 1mm
In addition to the case of very small $\mu$, $\lambda_{\rm HS}$ parameters, we note that larger values for these couplings may also bear interesting consequences. The existence of new scalar field tends to change the nature of the EW phase transition, making it to become the first order, and re-opening the door for the electroweak baryogenesis ~\cite{Kuzmin:1985mm,Curtin:2014jma, Kozaczuk:2019pet} if in addition to that new sources of $CP$-breaking are added into the theory. 

\subsubsection{Fermion (neutrino) Portal}
\label{sssec:fermion_portal}

The neutrino portal operates with one or several HNL states. The general form of the neutrino portal can be written as
\begin{equation}
 {\mathcal L}_{\rm fermion} = 
 {\mathcal L}_{\rm SM} + {\mathcal L}_{\rm DS} + \sum F_{\alpha I} (\overline{L}_{\alpha} H) N_I
\end{equation}
where the summation goes over the flavour of lepton doublets $L_{\alpha}$, and the number of available HNLs, $N_I$ . The $F_{\alpha I}$ are the corresponding Yukawa couplings.
The dark sector (DS) Lagrangian should include the mass terms for HNLs, that can be both Majorana or Dirac type.
Setting the Higgs field to its v.e.v., and diagonalizing mass terms for neutral fermions, one arrives at $\nu_i - N_J$ mixing, that is
usually parametrized by a matrix called $U$. Therefore, in order to obtain interactions of
HNLs, inside the SM interaction terms, one can replace $\nu_{\alpha} \to \sum_I U_{\alpha I} N_I$. In the minimal HNL models, both the production and decay of an HNL are controlled by the $U$ matrix elements. While it would be difficult to imagine a case where an HNL would couple to only one linear combination (or a single flavour) of charged leptons, it is nevertheless customary to assume a single flavour dominance as a way of comparing experimental reach. 

The emergence of $N$'s in weak charged and neutral currents immediately signals that the best way of producing and detecting the HNLs are via the decays of on-shell or off-shell $W$ and $Z$ bosons. Therefore, similar to the case of scalar portal, the LHC experiments, various flavour experiments, neutrino beam experiments and hadronic beam dumps are among the most promising venues in searches for HNLs. At the same time, the number of currently planned experiments that could deliver competitive sensitivity to $\{m_N, U_{\alpha}\}$ parameter space remains comparatively small.  

The cosmology and astrophysics of HNLs is very diverse, and we will not cover it in any substantial detail in this review, noting primarily that HNLs are a very promising DM candidate. If, for example, all $U_{\alpha I}$ leading to a keV-to-MeV mass HNLs are sufficiently small, they lead to an effective cosmological meta-stability. In that case, $\nu_{\rm SM}\to $HNL oscillations in the early Universe may lead to the correct cosmological abundance via freeze-in \cite{Dodelson:1993je}. However, the latest constraints on fluxes of X-rays resulting from HNL$_{\rm DM} \to \gamma \nu_{\rm SM}$ decays seem to indicate that the most straightforward mechanism for populating the DM is already excluded. Variants of this model {\em e.g.} with extra dark states interacting with HNLs may solve this under-abundance problem. 

\subsubsection{Pseudo-scalar Portal}
\label{sssec:pseudo_scalar_portal}

Taking a single pseudoscalar field $a$ one can write a set of its couplings to photons, gluons, and other fields of the SM. 
In principle, the set of possible couplings is very large and in
this study we consider only the flavour-diagonal subset,

\begin{equation}
\label{axion}
{\cal L}_{\rm axion} = {\cal L}_{\rm SM} + {\cal L}_{\rm DS} + \frac{a}{4f_\gamma} F_{\mu\nu} \tilde F_{\mu\nu} + \frac{a}{4f_G} {\rm Tr}G_{\mu\nu} \tilde G_{\mu\nu} +
\frac{\partial_\mu a} {f_l} \sum_\alpha  \bar l_\alpha \gamma_\mu\gamma_5 l_\alpha +
\frac{\partial_\mu a} {f_q} \sum_\beta  \bar q_\beta \gamma_\mu\gamma_5 q_\beta. 
\end{equation}

Due to the derivative nature of all perturbative interactions in (\ref{axion}), there is no direct feedback from quantum loops SM particles to the mass of $a$. (As already been mentioned, only $G\tilde G$ coupling can participate in contributing to $m_a$, at a nonperturbative level.) On the other hand, the power counting of divergencies in the perturbative renormalization of kinetic term
$(\partial_\mu a)^2$ shows quadratic sensitivity to UV scales, $\sim 
\Lambda^2f_a^{-2}$, signalling the need for UV completion. 
One can imagine that the DS Lagrangian may contain a new UV sector, the so-called the Peccei-Quinn sector,
with an approximate global symmetry, providing the required completion.
This is fundamentally different from vector, scalar and neutrino portals that do not require external UV completion, and requires in all the calculations the introduction of a cut-off scale. However, the laboratory reach to $a$ often has very weak dependence on details UV completion.

The cosmology of axion and axion-like particles is quite sensitive to its mass, coupling constant, and due to higher-dimensional nature of its coupling, may be sensitive to the initial conditions in the early Universe, such as reheat temperature after inflation. It is also important whether the Peccei-Quinn symmetry itself remained unbroken during inflation. Stringent limits on QCD axions can be derived from stellar energy loss arguments. Combined with estimates of cosmological abundance of axions starting from generic ({\em i.e.} not tuned) initial conditions, this creates an ``expectation band" for the mass-coupling relation that would give QCD axions the total energy density comparable to the observed DM, while astrophysical constraints typically favour sub-eV mass range. A more generic version - an ALP - can constitute DM for a much larger range of masses and couplings. 

Axion/ALP searches can be quite a versatile endeavour, especially if the mass range is extended to be above an MeV. They can be directly produced in colliders or beam dump experiments, or result from the FCNC decays of flavoured mesons. Similarly to the Higgs portal case, the generic axion coupling is not conserved, leading to sizeable loop-induced FCNC amplitudes. 
Experimental attention to ALPs above the MeV scale is relatively recent, with many new experimental and theoretical 
results to appear soon.

\subsection{Extending minimal portals}
\label{ssec:beyond_portals}

The framework reviewed in the previous Sections represents a guiding principle and captures many important phenomena in FIP physics. Yet the four portals may be extended to include some other important possibilities, some of which may have distinct observational signatures. Below we give a list of possible extensions to the four portal framework that are often considered, and enrich FIP phenomenology. Some of these extensions are driven by desire to accommodate some experimental anomalies (none of which so far has risen to the status of definitive). 
\begin{itemize}

\item[-] {\em Gravity portal.} All FIPs are guaranteed to have a gravitational coupling, and this is how the {\em e.g.} the existence of DM was inferred. There can be also additional couplings of new bosonic fields to gravity. For example, a new scalar field ($\varphi$) may have a direct coupling to the graviton ($h$) kinetic term via a dimension five operator, $M_*^{-1}\varphi \times (\partial h)^2$, leading to interesting particle phenomenology especially if $M_*\ll M_{Planck}$ and/or there are numerous fields $\varphi$ in existence, as in theories of large extra dimensions \cite{ArkaniHamed:1998rs}. 

\item[-] 
{\em Gauging of the SM accidental symmetries.}  
Dark photon portal with the $F'_{\mu\nu}B_{\mu\nu}$ coupling can be identically re-written in terms of a new gauge bosons $A'$ coupled to the same hypercharge quantum number as SM gauge boson $B_\mu$ \cite{Fayet:2006sp}, with $g_{B}\varepsilon$ coupling. It is then reasonable to explore the direct gauging of the SM accidental symmetries. The prime example of this is the  $U(1)_{B-L}$ gauge symmetry, often arising in the GUT and so-called ``left-right'' extensions of the SM gauge group \cite{Mohapatra:1980yp}. However, should the mass of the $B-L$ gauge boson be smaller than the EW scale, the coupling constant must be significanlty smaller than the SM couplings, $g_{B-L} \ll g_{\rm SM}$. A new and important experimental signature
tied to light $B-L$ boson is the modification to neutral current type neutrino scattering, with clear predictions for $\nu-e$ and $\nu-p(n)$
$B-L$ mediated amplitudes.

\item[-] {\em Flavour non-universal extensions.} Gauging individual flavour numbers in the quark sector may induce tree-level FCNC amplitudes, and come in strong conflict with flavour data. (Some of which, do in fact show slight deviations from the SM predictions \cite{Buttazzo:2017ixm}.) On the other hand, gauging some of the lepton numbers, such as $L_\mu-L_\tau$, does not lead to immediate flavour problems. Moreover, additional interactions, given to muons, may still be behind some of the discrepancies established in recent years, and in particular, $\sim 3\sigma$ tension with SM predictions in muon $g-2$ measurement. 

\item[-] {\em Extensions built  with $d\geq 6$ operators. }
Light FIPs may be attached to the SM by way of heavy mediators. Being heavy, these mediators do not have to be neutral under the SM gauge group. Integrating them out results in interactions that typically have dimensions $\geq 6$, suppressed by powers of high-energy scale. (Taking, for example, the dark photon mass to a large value reduces (\ref{int}) to a contact interaction $ \varepsilon m_{A'}^{-2} J_\mu^{\rm SM}J_\mu^{\rm dark}$.) The shear number of such operators that one can write down is very large, as it would have to include {\em e.g.} all different $J_\mu^{\rm SM}J_\mu^{\rm SM}$ combinations. The high-energy colliders, in general, provide the best possibilities for exploring such models, as the effect from such operators grows with energy. Studies of higher dimensional  operators built from the SM fields has been, of course, one of the major thrusts of particle physics over many decades. The possibility of accessing FIPs via higher dimensional    operators have come to focus relatively recently, in connection {\em e.g.} with ``hidden valley" \cite{Strassler:2006im} and ``neutral naturalness" ideas~\cite{Chacko:2005pe,Burdman:2006tz}, as well as in connection with dark matter searches at the LHC~\cite{Abercrombie:2015wmb}.

\item[-] {\em Neutron portal.} New FIP particles $\chi$ may be coupled to neutrons (or more generically to electrically neutral baryons) via $\bar n \chi+h.c.$ mixing terms, which are higher-dimensional operators, on account of the composite nature of $n$. If $m_\chi$ is at sub-nucleon level, novel decays of nucleons are possible, and in a relatively narrow mass window just below $m_n$, the long-lived nuclei are stable against decays to $\chi$ while neutron acquire novel decay channels \cite{Fornal:2018eol,Berezhiani:2018udo}. The new decay and $n-\chi$ oscillation channels are being explored at UCN facilities around the world. 

\end{itemize}

We saw that the portal framework allows one to have a  structured approach to physics of FIPs, and harmonise a large multitude of experimental searches. Yet this formalism 
can be considered as a gateway towards more fundamental and complete models, that hopefully can be better understood, once convincing FIP signals show up in experiments. 

In the following Sections we will mostly focus on results/projections of accelerator-based experiments obtained using the minimal portal framework, after having introduced in Section~\ref{ssec:astroparticle} how FIPs can be linked also to direct detection DM searches, astroparticle, and cosmology.

\subsubsection{Synergy and complementarity with direct DM detection experiments, astroparticle and cosmology}
\label{ssec:astroparticle}

\noindent
Rapid developments in DM direct detection have brought gains in sensitivity to WIMP cross sections, and importantly for our FIP discussion, lowered detection thresholds for some experiments. Consequently, some of the dark matter models discussed in connection with FIP portals may be probed in direct detection.

To make this connection well-defined, the spin choice and mass terms of the DM model must be specified. For example, in the absence of any $U(1)_{A'}$ breaking mass terms, complex scalar DM with a dark photon mediator (see Section~\ref{sssec:vector_portal}) predicts the scattering on electrons to be ($m_e\ll m_\chi$),
\begin{equation}
\label{DDscat}
    \sigma_{\chi e}  = \frac{16\pi \alpha \alpha_D\epsilon^2 m_e^2}{m_{A'}^4}= \frac{16\pi \alpha m_e^2}{m_\chi^4} \times y
    \simeq  3.7\times 10^{-27}\,{\rm cm}^2 \times (10\,{\rm MeV}/m_\chi)^4\times y,
\end{equation}
with a similar formula for the scattering on nuclei. Here $y$ is the same yield parameter that largely controls the abundance (see Eq.~\ref{eq:y}). Consequently, low threshold direct detection experiments probing DM-nucleus scattering can rule out significant fractions of parameter space for this model ({\em e.g.} $m_\chi > 400$\,MeV is disfavoured by CRESST (see \cite{Angloher:2015ewa} and discussion in Section~\ref{ssec:vector_exp}), while future planned DM-electron experiments have yet to achieve the levels of sensitivity relevant for sub-GeV freeze-out DM. 

\vskip 2mm
It is important to stress the underlying physical complementarity between direct detection and accelerator probes, especially when comparisons are made. Ultimately, direct detection probes DM deep in the non-relativistic limit, whereas accelerator experiments probe DM interactions in the relativistic limit, and this has profound consequences on the phenomenology. For example, different choices of DM spin can lead to direct detection rates suppressed by multiple powers of DM halo velocity $v\sim 10^{-3}$ among different models, whereas accelerator rates are only mildly affected by changing the spin because production occurs near $v\sim 1$. Additionally, direct detection rates are extremely sensitive to even small perturbations of the DM mass terms potentially present in ${\cal L}_{\rm mass}$ in Eq.~\ref{eq:L_vect}. In particular, the $U(1)_{A'}$ breaking mass term $\Delta m^2\chi^2$ generates a mass splitting $\Delta m$ between the real components of $\chi = 2^{-1/2}(\chi_1+i\chi_2)$ that is easily larger than the kinetic energy of WIMPs today ($E_{\rm kin} \sim \frac12 m_\chi c^2 (v_{\rm SM}/c)^2 \sim 10^{-5}{\rm eV}\times m_\chi/20\,{\rm MeV}$). In this case, $s$-wave elastic direct detection scattering is completely quenched, while having a negligible effect on the primordial abundance. Similar arguments apply to fermionic DM as well. Therefore, the direct detection program alone cannot be viewed as a ``universal substitute" to the accelerator-derived probes of the vector portal DM models. Likewise, accelerator probes alone cannot verify that the particles produced are indeed cosmologically long-lived. Both types of experiments are crucially important. 

\vskip 2mm
As already mentioned before, the early Universe puts serious restrictions on many variants of the portal models, and carves out significant parts of the parameter space, in what appears to be direct competition with accelerator-derived bounds. For example, the 100\,MeV-GeV scale HNLs are partly excluded by the primordial  nucleosynthesis (BBN). The production in the early Universe, even with initially absent HNLs results in their substantial presence during the neutron-proton interconvertion freeze-out. This typically limits the lifetimes of HNLs to a fraction of a second, resulting in tight limits on mixing angle versus mass. Similar considerations refer to the Higgs-mixed scalar. 
Dark photons in the experimentally accessible mixing angle range are typically too short-lived to be limited through BBN. On the other hand, if dark photons are supplied with WIMP-type dark matter, its annihilation to the electron-positron pairs is constrained. Too light dark matter, {\em i.e.} below a few MeV, has an unmistakable signature of injecting more energy to the photon fluid, therefore effectively reducing $N_{eff}$, a
commonly used parameter characterizing neutrino energy density relative to photons. For many other models, FIPs may lead to an increase in $N_{eff}$, such as for examples in models of milli-charged dark particles coupled to the SM via massless $A'$. 
In addition, as already mentioned, residual annihilation of DM during the recombination will lead to the discrepancy of the CMB angular anisotropy data with $\Lambda$CDM. This imposes strong restrictions on admissible WIMP models. However, declaring $s$-wave annihilating light WIMPs to be excluded would be an over-reach. Indeed, the same mass splitting perturbations in ${\cal L}_{\rm mass}$ may leave freeze-out annihilation unchanged, while significantly suppressing the annihilation at $z\sim 10^3$ and below, relevant for the CMB bounds. Again, one observes the complementarity of cosmological bounds to direct searches. 

Finally, stellar energy loss, and especially supernovae provide relevant bounds on MeV-100 MeV scale light particles. If too much energy is taken away from the exploding star, the energetics of the explosion changes, potentially affecting its development, and degrading the neutrino signal observed from the SN 1987a. As a result, similarly to the beam dump studies, SN exclude certain regions of parameter space, corresponding to sufficient production, and relatively small absorption/rescattering of newly produced particles inside the star.

\section{EXPERIMENTAL LANDSCAPE}
\label{sec:experiments}

\vskip 2mm
Searches for FIPs at accelerator based experiments have recently gotten a lot of attention in the experimental high-energy physics (HEP) community because they are the best placed to target the well motivated MeV-GeV mass range, as discussed in Section~\ref{sec:motivation}. Results from many past experiments have been reinterpreted within the portal formalism, new searches are being performed in many existing experiments at colliders and extracted beam lines, and new proposals are presented in almost all the main laboratories in the world.

\vskip 2mm
Accelerator experiments represent a unique tool to test models with light DM in the MeV-GeV range. The advantage of accelerator experiments is that the DM is produced in the relativistic regime, in a controlled manner, and at an energy scale similar to that which governs the annihilation and thermal decoupling of DM from the SM in the early Universe. This allows one to make predictions about the strength of accelerator DM signals based on the observed abundance of DM. Because the energy scales of thermal freeze-out and accelerator production are similar, these predictions tend to be fairly robust against variations of the underlying model that might severely impact non-relativistic signals such as scattering in direct detection (see Section~\ref{ssec:astroparticle}). 

\vskip 1mm 
Beyond light DM, FIPs with very small SM couplings are naturally very long-lived particles compared to SM particles, hence motivating experiments with long decay volumes to detect FIPs decays in visible final states. The presence of feeble couplings means also that these decays are very rare events, hence effective background rejection is paramount. To this aim, detectors need to have excellent tracking systems with the capability to precisely reconstruct displaced vertices and identify different species of hadrons and leptons for visible FIP decays. Another strategy is to precisely determine the kinematics of the events to constrain FIPs that are sufficiently long-lived to escape direct detection (as missing mass or missing momentum techniques, see Sections~\ref{ssec:FIPs-missing-p-E}-\ref{ssec:FIPs-missing-m}). 

\vskip 1mm
FIPs with mass in the MeV-several GeV range can be optimally searched for at fixed-target, beam dump, collider, and accelerator-based neutrino experiments using different techniques that depend on the characteristics of the available beam line. A general overview about the main experiments and future proposals searching for FIPs at a broad spectrum of experimental facilities is provided in Sections~\ref{ssec:FIPs-fixedtarget-beamdump},~\ref{ssec:FIPs-flavor-neutrino}, and~\ref{ssec:FIPs-LHC}. A description of the most commonly used experimental techniques is provided in Section~\ref{sec:experimental_techniques}. Finally Section~\ref{sec:exp_sensitivity} shows the multitude of results for FIP searches and projections for future experimental initiatives following the scheme of the minimal portal framework, as discussed in Section~\ref{ssec:portal_formalism}.

\subsection{FIPs searches at fixed-target and beam dump experiments}
\label{ssec:FIPs-fixedtarget-beamdump}
FIPs with masses in the MeV-GeV range can be produced in the interactions of a proton, electron, or muon beam with the material of a dump or an active target.
Given the feebleness of the couplings that drive FIP production and decays, high intensity beams and dedicated detectors are required to improve over the current results.

\vskip 2mm
 The {\it NA64$_e$} experiment~\cite{Banerjee:2016tad},  is searching for photon-mixed FIPs using $\sim 10^{11}$ electrons-on-target (eot) collected at the $\sim$100~GeV extracted electron beam line at the Super-Proton Synchrotron (SPS) at CERN.
 NA64$_e$ expects to increase by almost 10 times the data set in the close future (2021-2024) operating  with the same electron beam with increased intensity and higher energy an upgraded detector~{\it NA64$^{++}_e$}~\cite{NA64:eplus}. The NA64 collaboration plans to further expand its physics reach with the {\it NA64$_{\mu}$} detector~\cite{NA64:2018iqr}  searching for FIPs coupled to the second lepton generation. $NA64_{\mu}$ will be served by the 160~GeV extracted muon beam at the CERN SPS and will collect up to $\sim 10^{13}$~muons-on-target (mot) by 2024.
FIPs coupled to the second lepton generation can be searched for also by the {\it $M^3$} experiment~\cite{Kahn:2018cqs} proposed at a muon extracted beam line at the Fermi National Laboratory (FNAL) and possibly collecting a data set corresponding to $10^{10}$ ($10^{13}$)~mot in two subsequent phases of operation. 
Both experiments (NA64$_{\mu}$ and $M^3$) can search for feebly-interacting light new physics that could possibly explain the discrepancy of the measured $(g-2)_{\mu}$ value with the SM prediction.
 
\vskip 2mm
The {\it BDX} experiment~\cite{Battaglieri:2016ggd}, using the 11~GeV $e^-$ beam at the CEBAF facility at Jefferson Laboratory (JLAB), will search for light DM with $10^{22}$~eot 
by measuring the DM-$e^-$ scattering off a detector material downstream of a dump.
Similar technique can be possibly used by {\it DarkMESA} at Mainz~\cite{Christmann:2020qav} operating at the MESA facility~\cite{Doria:2019sux}, a continuous-wave high-intensity linac that will be able to provide an electron beam of  up to $155$~MeV energy. 
The fixed-target {\it HPS} experiment~\cite{Celentano:2014wya}, served by a multi-GeV beam line at the CEBAF facility at JLAB, can search for dark photon visible decays. After two short engineering runs in 2015 and 2016, 
HPS has taken two weeks of data in summer 2019 and four weeks are scheduled in 2021.  The {\it PADME} experiment~\cite{Raggi:2015gza} is currently taking data to search for dark photons, ALPs and dark scalars using the interactions of a 550~MeV positron beam with a diamond target at the Beam Test Facility (BTF) at Laboratori Nazionali di Frascati (INFN).

\vskip 2mm
CERN has the opportunity to play a leading role in these searches by fully exploiting the opportunities offered by the SPS with the proposed proton Beam Dump Facility (BDF)~\cite{Ahdida:2019ubf}, possibly serving the {\it SHiP} experiment~\cite{Anelli:2015pba} in the long term (2030$^{++}$) future. SLAC can be a leader in the search for light DM in the sub-GeV range with the recently approved S30XL beam line~\cite{Raubenheimer:2018mwt}  serving the proposed {\it LDMX} experiment~\cite{Akesson:2018vlm} with $10^{14}-10^{16}$~eot.
LDMX, in the two phases of operation with a 4 and 8~GeV electron beam, aims at covering the entire parameter space defined by a photon-mixed mediator in the MeV-GeV DM mass range compatible with the hypothesis that DM is a thermal relic from the early Universe. LDMX will be complemented by Belle II~\cite{Kou:2018nap} above the $\sim$ 1~GeV threshold. The possibility to use a 16~GeV electron beam at CERN~\cite{LDMX:CERN} would further extend the LDMX physics reach towards larger masses.

\subsection{FIPs searches at flavour, nuclear, and accelerator-based neutrino experiments}
\label{ssec:FIPs-flavor-neutrino}

Existing flavour-oriented experiments 
are exploiting a novel synergy between rare meson decays and exotic signatures of feebly-interacting particles and acting as main players in this emerging field.
In fact, as explained in Sections~\ref{sssec:scalar_portal} and \ref{sssec:pseudo_scalar_portal}  scalar and pseudo-scalar FIPs can be linked to the phenomenology of flavour-changing neutral currents bringing to enhanced $t-W$ loop-induced amplitudes in $b \to s$ and $s \to d$ transitions. 

{\it LHCb}~\cite{Aaij:2016qsm, Aaij:2015tna} and {\it Belle}~\cite{Wei:2009zv} searched for light dark scalars decaying to visible final states in the processes $B^+ \to K^+ \mu^+ \mu^-$ and $B^0 \to K^{*,0} \mu^+ \mu^-$, and obtained the strongest bound to date for light dark scalars in the mass range between the $\mu^+ \mu^-$ threshold and the kinematic limit given by $m_B - m_{K^{(*)}}$. These results are expected to be updated in the coming years by LHCb and Belle II.
Flavor experiments at $B$-factories, as {\it BaBar}~\cite{Lees:2017lec} in the past and  {\it Belle II}~\cite{Kou:2018nap} in the close future can also look for events with a single high-energy photon and large missing momentum and energy, consistent with production of a spin-1 light vector $A'$ particle through the process $e^+ e^- \to \gamma A'$, $A' \to $ invisible final states. The current BaBar result dominates the searches for light DM coupled with a light vector state above $\sim 1~$GeV threshold.

\vskip 2mm
A new interest towards FIPs in the low-mass range is also emerging in the current and future {\it accelerator-based neutrino experiments}, both in US
and Japan. 
In fact, the combination of high intensity proton beam facilities and large and massive detectors for precision neutrino parameter measurements makes short-baseline and near detectors of long-baseline neutrino experiments well suited for FIPs searches.

\vskip 2mm
{\it Kaon factories} and {\it neutrino experiments} are very well suited to search for HNLs coupled to the first two active neutrino generations produced in kaon decays. The {\it NA62} experiment~\cite{NA62:2017rwk} is searching for HNLs coupled to the first~\cite{NA62:2020mcv} and second\footnote{See talk by E. Goudzovski at KAON 2019 conference.} lepton generation in the decays $K^+ \to e^+ N$ and $K^+ \to \mu^+ N$,
reaching a sensitivity comparable to that published by the {\it T2K-ND280} near detector experiment~\cite{Abe:2019kgx}. 
Both experiments are expected to improve their results in the close future.
The reinterpretation of the analysis of $K^+ \to \pi^+ \nu \overline{\nu}$ decays~\cite{CortinaGil:2018fkc,CortinaGil:2020vlo} will allow the NA62 collaboration to reach an unprecedented sensitivity below the kaon mass for dark scalar or pseudo-scalar particles $X$ possibly produced in $K^+ \to \pi^+ X$ decays.
The NA62 experiment can expand the sensitivity to FIPs above the $K$ mass by operating the experiment in beam-dump mode~\cite{NA62:dump} and collecting about $10^{18}$ protons-on-target (pot) in dedicated periods of data taking during the 2021-2024 run.

\vskip 2mm
 The {\it MEG} experiment~\cite{Adam:2013vqa,TheMEG:2016wtm} (and in close future {\it MEG II}~\cite{Baldini:2013ke})
 at the Paul Scherrer Institute (PSI) is served by a high intensity ($10^8-10^{10}$ $\mu$/sec) low-momentum continuous muon beam that is perfectly suited to search for FIPs coupled to the second lepton generation. The experiment is searching for the charged lepton flavor violating (cLFV) muon decay $\mu \to e \gamma$. As a by-product of the main physics goal, MEG can search for a new light particle $X$  in the process $\mu \to X e, X \to \gamma \gamma$ ~\cite{Baldini:2020okg}.
 Also the Mu3e experiment~\cite{Berger:2014vba} at  PSI, whose main goal is to search for the cLFV $\mu^+ \to e^+ e^- e^+$ decay,  will search for  dark photon decays in the $e^+ e^-$ final state in the process $\mu^+ \to e^+ \nu_e \overline{\nu}_{\mu} A'$, $A' \to e^+ e^-$ ~\cite{Echenard:2014lma}. Charged pion decays and pion capture can also be used for searches of light FIPs \cite{Egli:1989vu}.
 
 \vskip 2mm
 A rather unique sensitivity to FIPs with masses below $\sim 20$\,MeV can be achieved using energy release in nuclear processes with a few nucleons. Old efforts in this direction \cite{Freedman:1984sd} were recently revived by the ATOMKI group \cite{Krasznahorkay:2015iga} with a 
 claim of experimental anomaly consistent with a 
 $\sim 17$\,MeV bosonic FIP. It is clear that this type of studies can be significantly advanced with the use of high-intensity low-energy beams at nuclear facilities. 

\vskip 2mm
{\it Short baseline neutrino experiments} at FNAL are expected to play an important role in the close future. {\it MicroBooNE}, {\it SBND} and {\it Icarus}~\cite{Antonello:2015lea} are all supposed to operate at the 8~GeV Booster beam line. 
MicroBooNE is already running, SBND and Icarus are expected to take data very soon. Being mostly focused to the detection to $\sim$~eV sterile neutrinos, these experiments are sensitive also to a variety of dark sector models. 
In the recent past the {\it MiniBooNE}~\cite{AguilarArevalo:2008qa} experiment, also  operated at the Booster neutrino beam at FNAL, presented  a new search for light DM performed with $10^{20}$ pot~\cite{Aguilar-Arevalo:2018wea}
and demonstrated that short-baseline neutrino experiments at proton beam dumps are an effective search method for DM with mass below 100~MeV.
On a longer time scale the {\it Multi-Purpose near Detector (MPD)} of the DUNE experiment~\cite{Berryman:2019dme} operating at the 120~GeV proton beam at the FNAL Main Injector will search for a multitude of light, feebly-interacting particles originated from  $\pi,K,D$ decays.

\subsection{FIPs searches at the LHC}
\label{ssec:FIPs-LHC}

FIPs with masses around the EW scale or above could be copiously produced in the high-energy $pp$ interactions at the LHC.
The LHC experiments are therefore the best placed to search for FIPs in the high mass range.
However, due to the feebleness of the couplings, FIPs can generically have lifetimes that are long compared to SM particles or the  majority of searches for new physics at the weak scale and can decay far from the interaction vertex of the primary proton-proton collision. That is why the sensitivity of the LHC existing detectors is typically limited in a region of relatively large couplings where lifetimes are typically short.
Moreover such signatures  often require customized techniques to reconstruct, for example, significantly displaced decay vertices, tracks with atypical properties, and short track segments. 

To overcome these limitations, a new set of dedicated detectors placed far apart from the LHC interaction points has been recently proposed at CERN -  
{\it MATHUSLA}~\cite{Alpigiani:2018fgd, Alpigiani:2020tva}, {\it FASER} (which is currently under construction), {\it FASER2}~\cite{Ariga:2018pin,Ariga:2018uku}, {\it CODEX-b}~\cite{Aielli:2019ivi}, {\it milliQan}~\cite{Ball:2016zrp}, {\it ANUBIS}~\cite{Bauer:2019vqk} and {\it MoeDAL-MAPP}~\cite{Frank:2019pgk} - that will be able to extend the physics reach for large masses in a range of feeble couplings inaccessible to the standard LHC experiments. 

\section{EXPERIMENTAL TECHNIQUES}
\label{sec:experimental_techniques}

Searches for FIPs at accelerator-based  experiments are performed with a large variety of beam lines and cover an impressive range of masses, from a few~MeV to a few~TeV.
Many different experimental techniques are used. They depend on the characteristics of the available beam lines and detectors and can be categorised as follows.

\subsection{Detection of visible decays}
\label{ssec:FIPs-visible}
HNLs, ALPs and in general light DM mediators feebly-coupled to SM particles can decay to visible final states with a probability that depends on the model and scenario. 
For example, if the mediators between SM and light DM particles have a mass which is less than twice the DM mass, once produced they can only decay to visible final states.

\vskip 2mm
The detection of visible final states is a technique used in almost all existing experiments searching for FIPs but it is mostly used by experiments operating with proton beams (either beam-dump and $pp$ colliders) where the background is typically very large and the kinematics of the signal production process is often not known. Typical signatures are expected to show up as narrow resonances over an irreducible background. The use of this technique requires high luminosity colliders or large fluxes of protons/electrons on a dump because the FIPs detectable rate is proportional to the fourth power of the coupling involved, $g^4$, and so very suppressed for very feeble couplings. Moreover visible FIPs decays to SM particles require  long decay volumes followed by spectrometers with excellent tracking systems and particle identification capabilities in order to disentangle very long-lived and elusive signals from usually large backgrounds.

\vskip 2mm
Existing or past experiments using this technique are: i) at $e^+e^-$ colliders: Belle~\cite{Abashian:2000cg}, BaBar~\cite{Aubert:2001tu}, KLOE~\cite{Adinolfi:2002uk, Adinolfi:2002zx}, and
Belle-II~\cite{Abe:2010gxa}; ii) at $pp$ colliders: ATLAS~\cite{Aad:2008zzm}, CMS~\cite{Chatrchyan:2008aa}, LHCb~\cite{Alves:2008zz}; iii) at proton beam dumps: CHARM~\cite{Winter:1982sk} at CERN; iv) at electron beam dumps: E137~\cite{Bjorken:1988as}, E141 at SLAC~\cite{Riordan:1987aw}; NA64$_e$~\cite{NA64-proposal} at CERN;
HPS~\cite{Celentano:2014wya} and APEX~\cite{Abrahamyan:2011gv} at JLAB.
Future proposals aiming to use this technique are: i) at the LHC: MATHUSLA~\cite{Alpigiani:2018fgd,Alpigiani:2020tva}, FASER(2)~\cite{Ariga:2018pin,Ariga:2018uku}, CODEX-b~\cite{Aielli:2019ivi}, ANUBIS~\cite{Bauer:2019vqk}, MoeDAL/MAPP~\cite{Frank:2019pgk}; ii) at proton beam dumps:  DarkQuest (or SeaQuest)~\cite{Aidala:2017ofy} at FNAL; NA62 in dump mode~\cite{NA62:dump} and SHiP~\cite{Anelli:2015pba} at CERN;
iii) at $e^+ e^-$ colliders: BES III~\cite{Ablikim:2019hff}.

\subsection{Direct detection of light DM scattering off the detector material}
\label{ssec:FIPs-DM}
Light DM produced in reactions of electrons and/or protons with a dump can travel across the dump and be detected via the scattering off the electrons and/or protons of the detector.
This technique has the advantage of probing directly the DM production processes but requires a large proton/electron yield to compensate the small scattering probability (as shown, for example, in Eq.~\ref{DDscat}).

Moreover the signature is very similar to that produced by neutrino interactions and this is a limiting factor for proton beam dump experiments unless it is possible to use a bunched beam and time-of-flight techniques. Of course this is much less a concern for electron beam dump experiments. Both for electron and proton beam dump experiments large factors can be gained by considering secondary products in the shower, as for example positrons annihilating with the electrons in the medium, as potential sources of light DM.

Existing or past experiments using this technique:
i) proton beam dumps: LSND~\cite{deNiverville:2011it},
MiniBooNE~\cite{AguilarArevalo:2008qa}, and MicroBooNE~\cite{Antonello:2015lea} at FNAL;
ii) electron beam dumps: BDX~\cite{Battaglieri:2014qoa} at JLAB.
Future experiments: SBND~\cite{McConkey:2017dsv} and MPD~\cite{Berryman:2019dme} at FNAL.

\subsection{Missing momentum/energy techniques}
\label{ssec:FIPs-missing-p-E}
Very-long lived FIPs  or FIPs decaying into light DM states can be detected in fixed-target reactions (as, for example,  $e^- Z \to e^- Z A'$, where $A'$ is a dark photon) by measuring the missing momentum ($\cancel{\it{p}}$) or missing energy ($\cancel{\it{E}}$) carried away from the escaping invisible particle(s). As such, these techniques apply to a rather broad class of models. 

\vskip 2mm
The missing momentum  technique exploits the change in the kinematics of a particle induced by crossing a target without being completely stopped, that can be measured by tracking systems upstream and downstream of the target. This technique will be mostly used by the proposed experiments LDMX~\cite{Akesson:2018vlm} at SLAC, $M^3$~\cite{Kahn:2018cqs} at FNAL, and NA64$_{\mu}$~\cite{NA64:2018iqr} at CERN.
In the case of missing energy, instead, the signal is a beam energy loss inside an active calorimeter without any other activity in the detector. The NA64$_e$ experiment at CERN~\cite{Banerjee:2016tad} is currently the only experiment exploiting this technique.

\vskip 2mm
For an electron beam experiment, the missing energy technique is challenged by charged-current neutrino production beyond $10^{14}$~eot, where the incoming electron converts into a neutrino and soft recoiling nucleus. On the other hand, missing momentum relies on a change in the transverse momentum of the electron before and after the production process, which is tagged. As such, charged-current backgrounds are mitigated (because the scattered electron is tagged), but neutral-current neutrino backgrounds are still a challenge beyond $10^{17}$~eot, depending on beam energy. Another attractive feature of transverse momentum is that it's sensitive to the mass of the underlying particles that are produced. 

\vskip 2mm
The main challenge of these two approaches is the high required detection efficiency for multi-GeV neutrons and kaons in order to veto events where a hard photon (from the incoming electron) converts into a few-body neutral hadronic final state. This requires detectors with excellent hermiticity in the forward direction, and high fidelity electromagnetic and hadronic calorimeter. However, these techniques have an intrinsic advantage in sensitivity for the same luminosity compared to techniques that require re-scattering or decay of the DM candidate/FIP, as they require neither re-scattering or decay, and therore have sensitivity that scales only as the SM-mediator coupling squared, $g^2$ or $\varepsilon^2$.

\subsection{Missing mass technique}
\label{ssec:FIPs-missing-m}
The missing mass ($\cancel{\it{M}}$) technique  is mostly used to detect invisible particles in reactions with a well-known initial state, as for example: i) at $e^+ e^-$ collider experiments (Belle~II and in the future BES~III~\cite{Ablikim:2019hff}) using the process $e^{+} e^{-} \to A' \gamma$, where a single photon is detected and nothing else;
ii) at kaon factories (NA62) via the decay $K\to \pi X$, where $X$ is either long-lived or decaying to light DM states;
iii) at positron fixed target experiment (PADME), via the process   $e^+ e^- \to A' \gamma$, where the positron annihilates with the electrons in the target and a single photon is detected in the final state and nothing else.

Characteristic signature is the presence of a narrow resonance 
emerging over a smooth background in the distribution of the missing mass. 
This technique requires detectors with very good hermeticity that allows the kinematics of the initial and final states to be fully controlled. 
The main limitation of this technique is the knowledge of the background arising from processes in which particles in the final state escape the apparatus without being detected.
 
 \vskip 2mm
\textbf{Table~\ref{tab:exps}} summarises the main  past, current, and future experiments aiming at detecting feebly-interacting particles, the beam type, the collected or expected data set, the experimental technique used, and the main portals the experiments are sensitive to. 

\clearpage
\begin{table}[h]
\tabcolsep7.5pt
\caption{Main past, current, and future accelerator-based experiments sensitive to FIPs searches. Legend for portals: 1: Vector; 2: Scalar; 3: Pseudo-scalar; 4: Fermion.}
\label{tab:exps}
\begin{center}
\begin{tabular}{|l|l|c|c|c|c|}
\hline 
Experiment & lab  & beam & particle yield/$\mathcal{L}$ & technique & portals \\
\hline
{\bf past} &     & & &  &  \\ \hline
{\scriptsize LSND~\cite{deNiverville:2011it}}     & {\scriptsize LANL}   &  {\scriptsize $p$, 800~MeV} & {\scriptsize $10^{23}$~pot} & {\scriptsize $e^-$ recoil} & (1) \\

{\scriptsize E137~\cite{Bjorken:1988as}}     & {\scriptsize SLAC}   & {\scriptsize $e^-$, 20~GeV} & {\scriptsize $2\cdot 10^{20}$ (30~C)} & {\scriptsize visible} & {\scriptsize (1,3)} \\ 

{\scriptsize E141~\cite{Riordan:1987aw}}     & {\scriptsize SLAC}  &  {\scriptsize $e^-$, 9~GeV} & {\scriptsize $2 \cdot 10^{15}$} & {\scriptsize visible} & {\scriptsize (1,3)}  \\ 

{\scriptsize E774~\cite{Bross:1989mp}}     & {\scriptsize FNAL}  &  {\scriptsize $e^-$, 275~GeV} & {\scriptsize $2\cdot 10^{15} $ } & {\scriptsize visible} & {\scriptsize (1)}  \\

{\scriptsize NuCAL~\cite{Blumlein:2011mv,Blumlein:2013cua}}    & {\scriptsize Serpukhov}     & {\scriptsize $p$, 70~GeV} & {\scriptsize $1.7 \cdot 10^{18} $} & {\scriptsize visibile} & {\scriptsize (1,3)} \\
{\scriptsize CHARM~\cite{Winter:1982sk}}    & {\scriptsize CERN}  &  {\scriptsize $p$, 400~GeV} & {\scriptsize $2.4~\cdot~10^{18}$} & {\scriptsize visible} & {\scriptsize (1,2,3,4)}  \\ 

{\scriptsize MiniBooNE~\cite{AguilarArevalo:2008qa}} & {\scriptsize FNAL}  & {\scriptsize $p$, 8~GeV} & {\scriptsize $1.9 \cdot 10^{20}$} & {\scriptsize recoil $e,N$} &  {\scriptsize (1)} \\
{\scriptsize KLOE~\cite{Adinolfi:2002uk,Adinolfi:2002zx}}     & {\scriptsize LNF}   & {\scriptsize $e^+ e^-$, 1~GeV} & {\scriptsize up to 1.7~fb$^{-1}$} &{\scriptsize visible, inv.} & {\scriptsize (1)} \\ 
{\scriptsize NA48/2~\cite{Fanti:2007vi} }  & {\scriptsize CERN}   & {\scriptsize $\pi^0$} & {\scriptsize $2 \cdot 10^7$} & {\scriptsize $\cancel{\it{M}}$} & {\scriptsize (1)}  \\ 

{\scriptsize Belle~\cite{Abashian:2000cg}}    & {\scriptsize KEK}    & {\scriptsize $e^+ e^-$, 10.58~GeV} & 0.6-0.8~fb$^{-1}$ &{\scriptsize visible} &{\scriptsize (1,2,4) }  \\ 

{\scriptsize BaBar~\cite{Aubert:2001tu}} & {\scriptsize SLAC}   & {\scriptsize $e^+ e^-$, 10.58~GeV} & {\scriptsize 514~fb$^{-1}$} & {\scriptsize visible, invis.}  & {\scriptsize (1) }\\  \hline

{\bf current}  &       & & &  &  \\ \hline
{\scriptsize ATLAS~\cite{Aad:2008zzm}}  & {\scriptsize CERN}  &  {\scriptsize $pp$, 13-14~TeV} & {\scriptsize up to 3~ab$^{-1}$} & {\scriptsize visible, invis.} & {\scriptsize (1,2,3,4) }\\

{\scriptsize CMS~\cite{Chatrchyan:2008aa}}   & {\scriptsize CERN} & {\scriptsize $pp$, 13-14~TeV} & {\scriptsize up to 3~ab$^{-1}$} & {\scriptsize visible, invis.} & {\scriptsize (1,2,3,4)} \\

{\scriptsize LHCb~\cite{Alves:2008zz}}  & {\scriptsize LHC} & {\scriptsize $pp$, 13-14~TeV}  & {\scriptsize up to 300~fb$^{-1}$} & {\scriptsize visible}  & {\scriptsize (1,2,3,4)} \\

{\scriptsize Belle II~\cite{Abe:2010gxa}} & {\scriptsize KEK} & {\scriptsize $e^+ e^-$, 11~GeV} & {\scriptsize up to 50~ab$^{-1}$ } & {\scriptsize visible, invis.} & {\scriptsize (1,2,3,4)}  \\

{\scriptsize BES III~\cite{Ablikim:2019hff} } & {\scriptsize BEPCII }  & {\scriptsize $e^+ e^-$, 3.7~GeV} & {\scriptsize up to 40~fb$^{-1}$} & {\scriptsize invis.} &  {\scriptsize (1)}  \\

{\scriptsize NA62~\cite{NA62:2017rwk}}   & {\scriptsize CERN}   &{\scriptsize $K^+$, 75}~GeV & {\scriptsize up to $10^{13}$ K decays} &  {\scriptsize visible, invis.} & (1,2,3,4) \\

{\scriptsize NA64$^{(++)}_{e}$~\cite{NA64:eplus}} & {\scriptsize CERN }   & {\scriptsize $e^-$, 100~GeV} & {\scriptsize up to $3\cdot~10^{12}$~eot/year} & {\scriptsize $\cancel{\it{E}}$, visible} & (1,3) \\

{\scriptsize HPS~\cite{Celentano:2014wya} } &    {\scriptsize JLAB}    & {\scriptsize $e^-$, 2-6~GeV} & {\scriptsize $\sim 10^{20}$~eot} & {\scriptsize visible} & {\scriptsize (1,3)} \\

{\scriptsize APEX~\cite{Abrahamyan:2011gv}}  & {\scriptsize JLAB} & {\scriptsize $e^+$, 2.2~GeV} & {\scriptsize up to 150~$\mu$A} & {\scriptsize visible} & {\scriptsize (1) } \\

{\scriptsize MicroBooNE~\cite{Antonello:2015lea}}  & {\scriptsize FNAL}   & {\scriptsize $p$, 8~GeV} & {\scriptsize $\sim 10^{21}$~pot}  & {\scriptsize recoil, Ar} &  (1) \\

{\scriptsize Dark(Sea)Quest~\cite{Berlin:2018pwi}} & {\scriptsize FNAL}   & {\scriptsize $p$, 120~GeV} & {\scriptsize $10^{18} \to 10^{20}$} & {\scriptsize visible }
& {\scriptsize (1,2,3,4)}  \\

{\scriptsize T2K-ND280~\cite{Abe:2019whr}} & {\scriptsize JPARC} &  {\scriptsize $p$, 30~GeV} &   {\scriptsize $10^{21}$~pot}  &  {\scriptsize visible}  &  {\scriptsize (4) }\\

{\scriptsize PADME~\cite{Raggi:2015gza}}  & {\scriptsize LNF} & {\scriptsize $e^+$, 550~MeV} & {\scriptsize $5 \cdot 10^{12}$ $e^+$ot} & {\scriptsize missing mass}  & {\scriptsize (1)} \\ 

{\scriptsize PIENU~\cite{Malbrunot:2011zz}}  & {\scriptsize TRIUMF}   &{\scriptsize $\pi^+$, 75~MeV}   & {\scriptsize $10^7$}  & {\scriptsize missing mass} & {\scriptsize (4)} 

 \\\hline

{\bf future/proposed} &            & & & & \\ \hline
{\scriptsize SHiP~\cite{Anelli:2015pba}}   & {\scriptsize CERN}   & {\scriptsize $p, 400$~GeV}  & {\scriptsize $2\cdot 10^{20}$/5 years} & {\scriptsize visible, recoil}   & {\scriptsize (1,2,3,4)} \\ 

{\scriptsize NA62-dump~\cite{NA62:dump}}  & {\scriptsize CERN}   & {\scriptsize $p$, 400~GeV}  & {\scriptsize $\sim 10^{18}$~pot}  & {\scriptsize visible}   & {\scriptsize (1,2,3,4)}  \\ 

{\scriptsize NA64$_{\mu}$~\cite{NA64:2018iqr}} & {\scriptsize CERN}  & {\scriptsize $\mu$, 160~GeV} & {\scriptsize up to $10^{13}$~mot/year} & {\scriptsize $\cancel{\it{p}}$} & {\scriptsize (1)} \\
{\scriptsize RedTop~\cite{Gatto:2019dhj}}   &  {\scriptsize CERN }  & {\scriptsize $p$, 1.8,3.5~GeV} & {\scriptsize up to $10^{17}$~pot} & {\scriptsize visible} & {\scriptsize (2,3)}  \\
{\scriptsize Dark MESA~\cite{Christmann:2020qav}}  & {\scriptsize Mainz}     & {\scriptsize $e^-$, 155~MeV} & {\scriptsize 150~$\mu$A}  & {\scriptsize visible} & {\scriptsize (1)} \\

{\scriptsize LDMX~\cite{Akesson:2018vlm}}  & {\scriptsize SLAC}        & {\scriptsize $e^-$, 4,8~GeV}  & {\scriptsize $2 \cdot 10^{16}$~eot} & {\scriptsize $\cancel{\it{p}}$, visible} & {\scriptsize (1)} \\ 

{\scriptsize SBND~\cite{McConkey:2017dsv}}  & {\scriptsize FNAL}   &{\scriptsize $p$, 8~GeV}     & {\scriptsize $6 \cdot 10^{20}$~pot} & {\scriptsize recoil Ar}  & (1) \\
{\scriptsize LBND~\cite{Berryman:2019dme}} &  {\scriptsize FNAL}     & {\scriptsize $p$, 120~GeV}   & {\scriptsize $\sim 10^{21}$~pot}    & {\scriptsize recoil $e,N$} 
& {\scriptsize (1,2,3,4)} \\

{\scriptsize M$^3$~\cite{Kahn:2018cqs}} & {\scriptsize FNAL} & {\scriptsize $\mu$, 15~GeV} & {\scriptsize $10^{10}$ ($10^{13})$}~mot & $\cancel{\it{p}}$ & (1)\\

{\scriptsize FASER(2)~\cite{Feng:2017uoz}}  & {\scriptsize CERN}  & {\scriptsize $pp$, 14~TeV} & {\scriptsize 150~fb$^{-1} \to 3$~ab$^{-1}$} & {\scriptsize visible}  & {\scriptsize (1,2,3,4)} \\ 

{\scriptsize CODEX-b~\cite{Aielli:2019ivi}} & {\scriptsize CERN}     &{\scriptsize $pp$, 14~TeV}   & {\scriptsize 300~fb$^{-1}$ }       & {\scriptsize visible}     & 
{\scriptsize (1,2,3,4)} \\ 
{\scriptsize MATHUSLA~\cite{Alpigiani:2018fgd}} & {\scriptsize CERN }  & {\scriptsize $pp$, 14~TeV}   & {\scriptsize 3~ab$^{-1}$ } & {\scriptsize visible}    & 
{\scriptsize (1,2,3,4)} \\ 
{\scriptsize milliQan~\cite{Ball:2016zrp} } & {\scriptsize CERN} &   {\scriptsize $pp$, 14~TeV}  & {\scriptsize 0.3-3~ab$^{-1}$} & {\scriptsize visible} & {\scriptsize (1)} \\
{\scriptsize MoeDAL/MAPP~\cite{Frank:2019pgk}}  & {\scriptsize CERN} & {\scriptsize $pp$, 14~TeV} & 
 {\scriptsize 30~fb$^{-1}$} & {\scriptsize visible} & {\scriptsize (4)} \\
 
{\scriptsize BDX~\cite{Battaglieri:2014qoa}}  &   {\scriptsize JLAB}   & {\scriptsize $e^-$, 11~GeV} & {\scriptsize $\sim 10^{22}$} & {\scriptsize recoil $e$} 
& {\scriptsize (1,3)} \\
{\scriptsize DarkLight~\cite{Balewski:2014pxa}} & {\scriptsize JLAB} & {\scriptsize $e^-$, 100 MeV} &{\scriptsize 5~mA} & {\scriptsize visible} & {\scriptsize (1) } \\ 

{\scriptsize Mu3e~\cite{Arndt:2020obb}}   & {\scriptsize PSI}  &  29~GeV  & {\scriptsize  $10^{18} \to 10^{20} \mu$/s} & {\scriptsize visible} & {\scriptsize (1)} \\

\hline
\end{tabular}
\end{center}
\end{table}

\clearpage
\section{EXPERIMENTAL SENSITIVITY}
\label{sec:exp_sensitivity}

\textbf{Table~\ref{tab:params}} summarises the parameters each portal is sensitive  to and corresponding signatures.
In the following Sections, for each portal, experimental bounds from past and current searches are compared against projections of existing or proposed experiments and bounds from several astrophysical and cosmological sources, computed within the same model (see Section~\ref{ssec:astroparticle}). 
It is important to note that astrophysical and cosmological probes tend to be less robust against modifications to the model than accelerator-based results, that instead represent a solid ground for bench-marking theoretical hypotheses.
It is also important to remark that the level of maturity in evaluating sensitivity projections is highly non homogeneous among the proposals and should be taken with many caveats. 

\begin{table}[h]
\tabcolsep7.5pt
\caption{Main parameters each portal is sensitive to and relative signatures.}
\label{tab:params}
\begin{center}
\vskip -2mm
\begin{tabular}{|c|c|c|}
\hline 
Portal & parameter space & signature \\ \hline
Dark Photon, $A'$ & $y = \alpha_D \varepsilon^2 \alpha (m_{\chi} / m_{A'})^4$ vs $m_{\chi}$ &  {\footnotesize DM scattering, $A' \rightarrow $ invisible } \\
     & $Q_{\chi}/e$    & {\footnotesize $A' \to $ millicharged fermions} \\
     & $ \varepsilon$ vs $m_{A'}$ & $A' \to $ visible modes  \\ \hline
Dark Scalar, $S$  & $\sin \theta$ vs $m_S$  & $S \to $  visible/invisible modes  \\ \hline

Dark Pseudo-Scalar, $a$ & photons: $g_{a \gamma\gamma}$ vs $m_a$ &  $a \to $ visible/invisible modes \\
                        & fermions: $g_{Y} = 2v/f_a$ vs $m_a$  & $a \to $  visible/invisible modes \\
                        & gluons: $ g_G = 1/f_G$ vs $m_a$   & $a \to $ visible \\ \hline
Heavy Neutral Lepton, $N$ & $U^2_e, U^2_{\mu}, U^2_{\tau}$ vs $m_N$ & $N \to $ visible/invisible modes \\ \hline \hline
\end{tabular}
\end{center}
\end{table}

\vskip -4mm
\subsection{Vector Portal: search for dark photons and light DM}
\label{ssec:vector_exp}

Motivated by models of light DM, much theoretical and experimental effort over the past decade has focused on the vector portal, which at low energies involves kinetic mixing between the photon and the dark vector, $\varepsilon/2  F^{\mu \nu}F^{'}_{\mu \nu}$.
This is in part because it is the least constrained scenario that allows for bilinear mixing. It also provides the most scope for model building, including what has become the benchmark model for sub-GeV DM 
(see, for example, Refs.~\cite{Alexander:2016aln,Battaglieri:2017aum, Beacham:2019nyx}).

As explained in Section~\ref{sssec:vector_portal}, the vector portal allows a minimally coupled viable WIMP-like DM model to be built with a light, vector mediator (dark photon, $A'$) and a DM candidate $\chi$.
In general, the usual analysis of accelerator-based searches starts from the assumption that the mediator is light enough to be produced on-shell, though off-shell production reactions are also important. 
The production processes of $A'$ depend on the available beam: it can be bremsstrahlung for an electron beam, annihilation with the electrons of a target for a positron beam, and meson decays, bremsstrahlung, and deep-inelastic scattering for a proton beam line.
Experimental signatures depend on the ratio between the vector mediator and dark matter particle masses,
$m_{A'} / m_{\chi}$. They can be: i) for $m_{A'}/m_{\chi} \geq 2$ an invisible decay of $A'$, $A' \to \chi \overline{\chi}$,
that can be detected via missing energy, missing momentum, and missing mass techniques or via DM scattering off the detector material; ii) for $m_{A'}/m_{\chi} < 2$, a visible decay of $A'$ into SM final states, and direct production of $\chi \overline{\chi}$ through an off-shell $A'$, which can again be detected with missing mass/energy techniques. 

\subsubsection{Light Dark Matter Production}
\label{ssec:vector_exp_ldm}
\begin{figure}[h]
\includegraphics[width = 11cm] {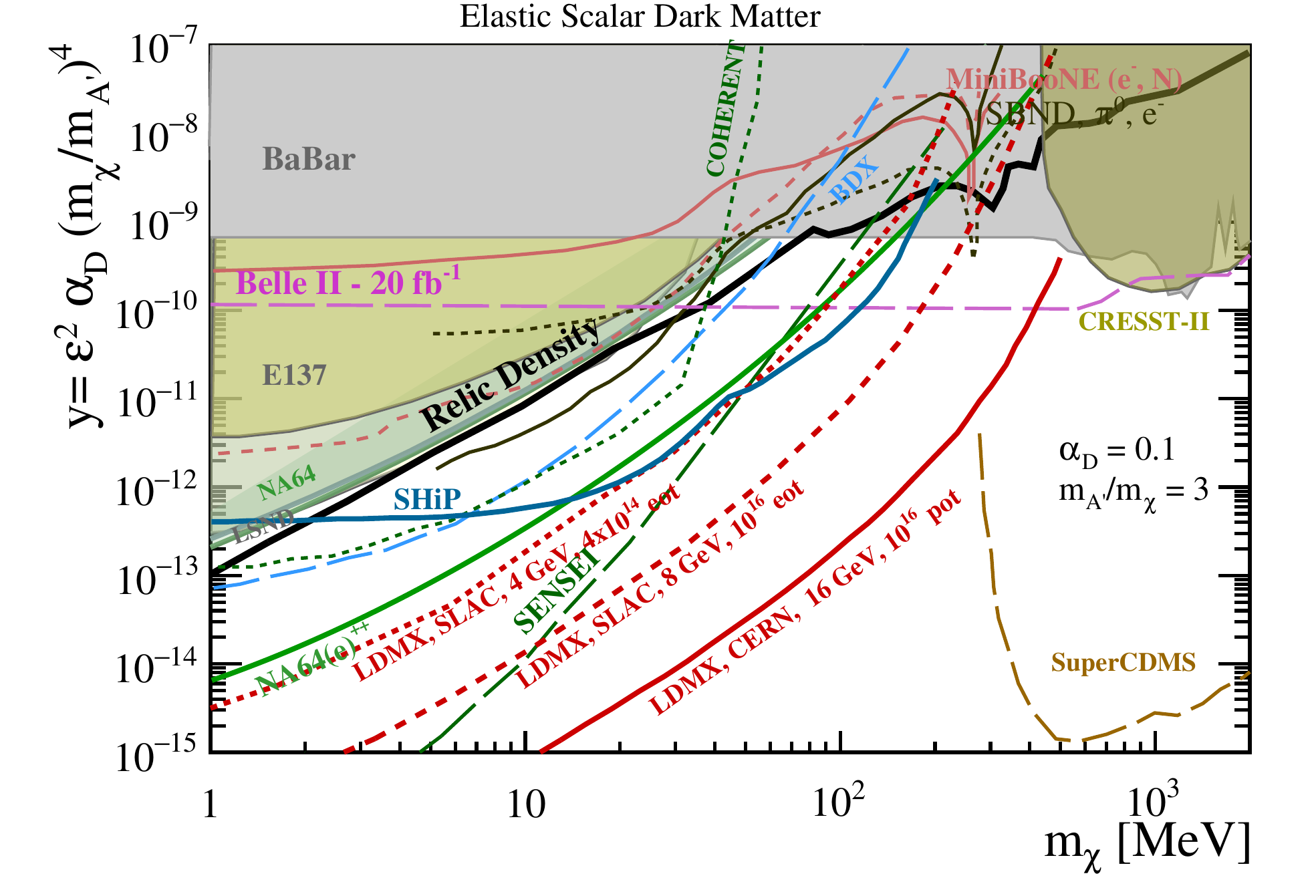}
\includegraphics[width = 11cm] {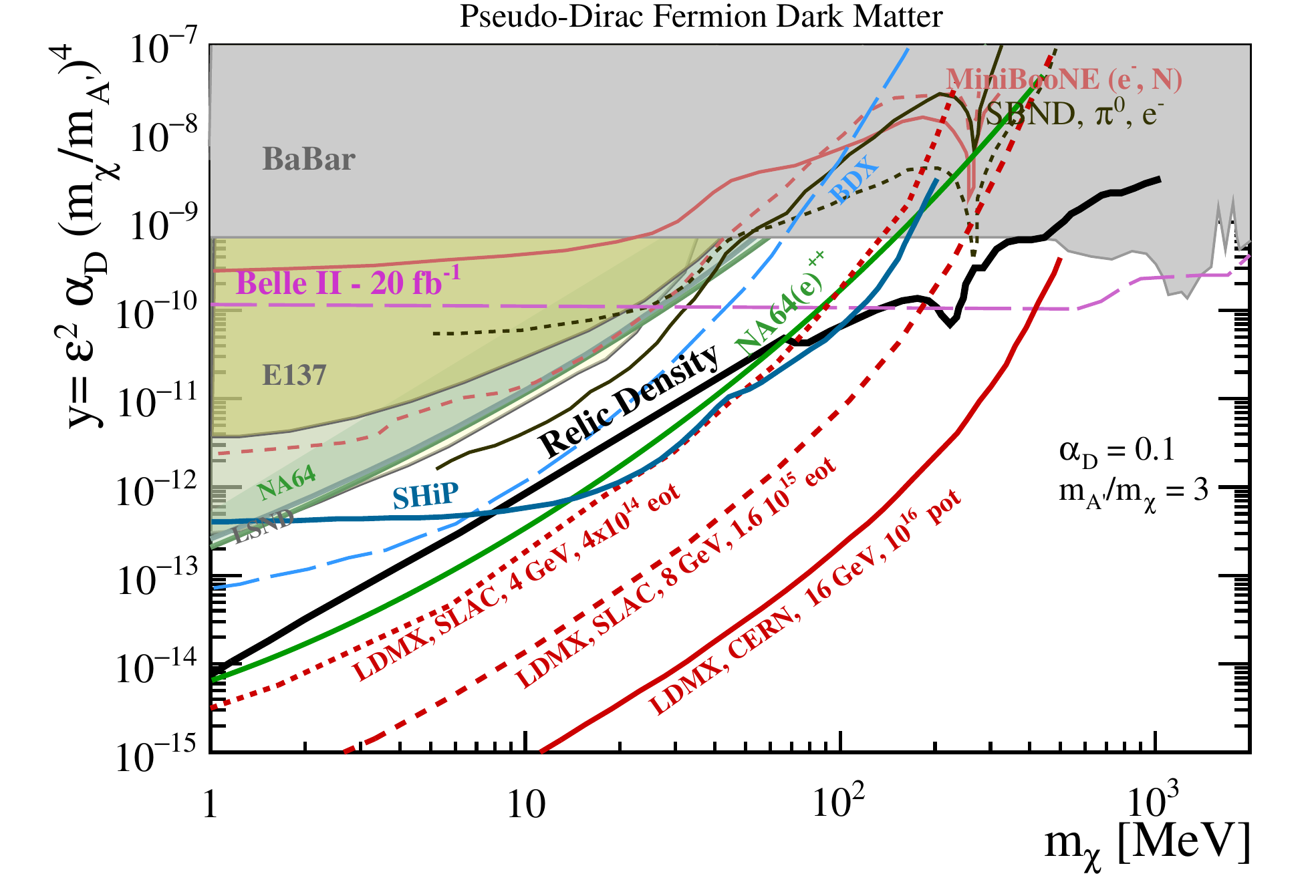}
\caption{\small 
  \label{fig:massive4}  
  Existing limits ( filled areas) and future sensitivities of existing or proposed experiments (coloured curves) to light dark matter production through a dark photon in the plane defined by the yield variable $y$ as a function of DM mass $m_{\chi}$ for a specific choice of $\alpha_D = 0.1$ and $m_{A'}/m_{\chi} = 3$. The DM candidate is assumed to be an elastic scalar (top) or a pseudo-Dirac fermion (bottom).  Limits shown as filled areas are: BaBar~\cite{Lees:2017lec}; NA64$_e$~\cite{NA64:2019imj}; reinterpretation of the data from E137~\cite{Batell:2014mga} and LSND~\cite{deNiverville:2011it};  result from MiniBooNE~\cite{Aguilar-Arevalo:2018wea}; inferred sensitivity from the direct detection DM experiment CRESST-II~\cite{Angloher:2015ewa}. The projected sensitivities, shown as solid or dashed lines, come from: SHiP~\cite{Anelli:2015pba}, BDX~\cite{Battaglieri:2016ggd}, SBND~\cite{Antonello:2015lea}, COHERENT~\cite{deNiverville:2015mwa}, LDMX@SLAC~\cite{Akesson:2018vlm,Raubenheimer:2018mwt}, LDMX@CERN~\cite{LDMX:CERN}, Belle-II~\cite{Kou:2018nap}; SENSEI~\cite{Battaglieri:2017aum} and SuperCDMS~\cite{Agnese:2016cpb}.}
 
\label{fig:DP_y_scalar}
\end{figure}

With the experimental effort to fully explore light DM now becoming a reality, it is timely to analyze the full thermal relic parameter space, in order to determine what capabilities are needed to fully test and either discover or exclude simple models of light thermal DM.
\textbf{Figure~\ref{fig:DP_y_scalar}} shows the lively activity of past, current, and proposed experiments to search for light DM with mass in the MeV-GeV range at accelerator-based experiments in all major laboratories in the world (CERN, KEK, FNAL, SLAC, and JLAB).
Results are compared against the current bounds for DM from cosmology under the hypothesis that the light DM candidate
is a scalar (top plot) and pseudo-Dirac fermion particle (bottom plot).

As explained in Section~\ref{ssec:astroparticle}, for a specific choice of dark matter particle $\chi$ (i.e. scalar, Majorana, pseudo-Dirac), accelerator-based experiments can be directly compared against a reinterpretation of results from DM direct detection experiments. But importantly, the quantitative results of this comparison depend strongly on the spin and other details of the DM particle. As discussed earlier, the origin for this dependence is connected to the underlying physical reason why accelerator and direct detection techniques are fundamentally complementary -- direct detection probes DM deep in the non-relativistic limit, whereas accelerator experiments probe DM interactions in the relativistic limit. As a result, different choices of DM spin can lead to direct detection rates suppressed by multiple powers of DM halo velocity $v\sim 10^{-3}$ among different models, whereas accelerator rates are only mildly affected by changing the spin. Thus, to give a representative view of the complementary nature of these techniques, we show two viable scenarios where the DM spin is different, namely a scalar and a pseudo-Dirac fermion (we don't show the Majorana fermion case). 

The current bounds are expressed in the parameter space \{$y, m_{\chi}$\} and come from: 
BaBar~\cite{Lees:2017lec}, NA64$_e$~\cite{NA64:2019imj}, reinterpretation of the data from old beam dump experiments as E137~\cite{Batell:2014mga} and LSND~\cite{deNiverville:2011it},  MiniBooNE~\cite{Aguilar-Arevalo:2018wea}, and - only in the case of scalar DM - interpretation of data from  the direct detection experiment CRESST-II~\cite{Angloher:2015ewa}. The projected sensitivities come from SHiP~\cite{Anelli:2015pba}, BDX~\cite{Battaglieri:2016ggd},  SBND~\cite{Antonello:2015lea},
Belle-II~\cite{Kou:2018nap},
LDMX@SLAC~\cite{Akesson:2018vlm,Raubenheimer:2018mwt},
LDMX@CERN~\cite{LDMX:CERN}, and again only in the case of scalar DM, some direct detection experiments as:
COHERENT~\cite{deNiverville:2015mwa}, SENSEI~\cite{Battaglieri:2017aum} with a proposed 100~g detector and SuperCDMS~\cite{Agnese:2016cpb} direct detection DM experiments, both expected to be operated at SNOLAB.

The sensitivity plots have been obtained for  $m_{A^{\prime}}/m_{\chi}= 3$ and of the coupling with DM $\alpha_D = 0.1$, which are quite commonly used in literature (though $\alpha_D = 0.5$ is the other common choice). The dependence of the bounds with these two parameters have been extensively studied in~\cite{Berlin:2020uwy}. While the dependence of the $y$ variable with $\alpha_D$  is implicit in the $y$ definition,
the dependence with $m_{A^{\prime}}/m_{\chi}$ is more involved.
However beyond the resonant process occurring at $m_{A^{\prime}} = 2 m_{\chi}$, the dependence of $y$ is a decreasing function with the mass ratio  and therefore the choice $m_{A^{\prime}} = 3 m_{\chi}$
generally yields a conservative estimate of experimental sensitivity within the region where the effective field theory is approximately valid at freeze-out. 
For this conservative choice of $\alpha_D$ and $m_{A^{\prime}}/m_{\chi}$ parameters, the region of parameter space corresponding to thermal freeze-out is close to saturated by existing data or will be in the very close future (Figure~\ref{fig:DP_y_scalar}, top)
if the DM is an elastic scalar particle.
In the case of a pseudo-Dirac fermion DM  (Figure~\ref{fig:DP_y_scalar} (bottom)) the thermal relic target is still far from the actual experimental bounds but can be reached in a not too far future (similarly for Majorana DM). The sensitivity to direct DM production through off-shell mediators -- which dominates accelerator production when $m_{A^{\prime}}$ is less than $2 m_{\chi}$ -- is considered in~\cite{Berlin:2020uwy}, but the conclusion with respect to future experimental coverage of thermal freeze-out targets remains the same. 

\clearpage
\subsubsection{Dark photon decays to visible final states}
\label{ssec:vector_exp_vis}
\begin{figure}[h]
\includegraphics[width=12cm]{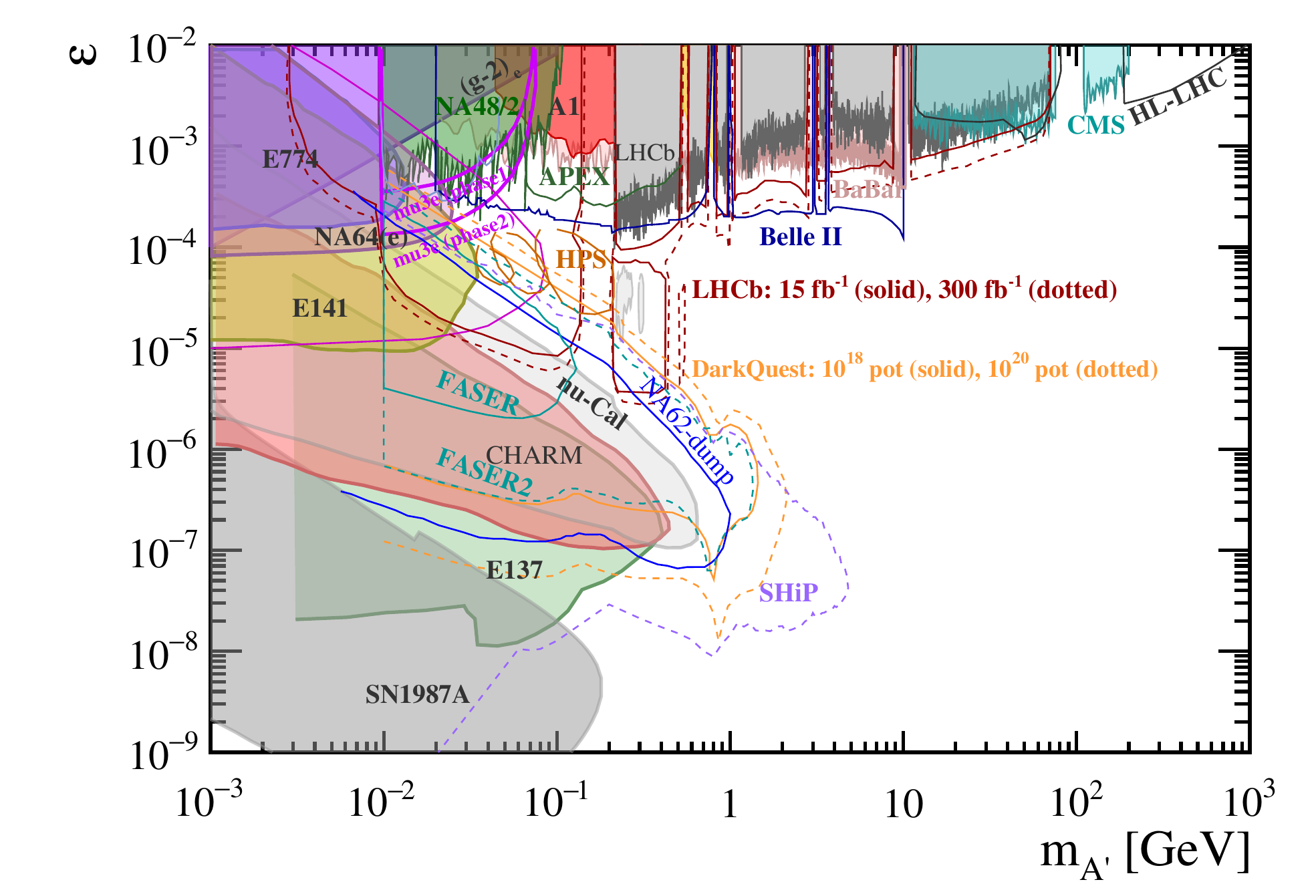}
\caption{ {\bf Dark photon into visible final states:} $\varepsilon$ versus $m_{A'}$. 
Filled areas are 
existing limits from 
searches at experiments at collider/fixed target (A1~\cite{Merkel:2014avp}, LHCb~\cite{Aaij:2019bvg},
CMS~\cite{CMS:2019kiy},
BaBar~\cite{Lees:2014xha}, KLOE~\cite{Archilli:2011zc,Babusci:2012cr,Babusci:2014sta,Anastasi:2016ktq}, and NA48/2~\cite{Batley:2015lha}) and old  beam dump:  E774~\cite{Bross:1989mp}, E141~\cite{Riordan:1987aw}, E137~\cite{Bjorken:1988as,Batell:2014mga,Marsicano:2018krp}), $\nu$-Cal~\cite{Blumlein:2011mv,Blumlein:2013cua},
and CHARM (from~\cite{Gninenko:2012eq}).
Bounds from supernovae~\cite{Chang:2016ntp} and $(g-2)_e$~\cite{Pospelov:2008zw} are also included.
Coloured curves are projections for existing and proposed experiments:
Belle-II~\cite{Kou:2018nap};
LHCb upgrade~\cite{Ilten:2016tkc,Ilten:2015hya}; NA62 in dump mode~\cite{NA62:dump} and
NA64(e)$^{++}$~\cite{NA64:eplus}; FASER and FASER2~\cite{Feng:2017uoz};
seaQUEST~\cite{Berlin:2018pwi}; HPS~\cite{Adrian:2018scb};  
Dark MESA~\cite{Doria:2019sux},
Mu3e~\cite{Echenard:2014lma}, and  HL-LHC~\cite{Curtin:2014cca}. Figure revised from Ref.~\cite{Fabbrichesi:2020wbt}.
}
\label{fig:DP_visible}
\end{figure}

If the DM is heavier than $m_{A'}/2$ or contained in a different sector, the $A'$ once produced, decays back to SM particles via the $\varepsilon-$proportional interactions between dark photon and SM particles. This is shown in \textbf{Figure~\ref{fig:DP_visible}}. 
Existing limits on the massive dark photon for $m_{A^{\prime}} >1$ MeV come from peak searches at experiments at collider/fixed target (A1~\cite{Merkel:2014avp}, LHCb~\cite{Aaij:2019bvg},
CMS~\cite{CMS:2019kiy},
BaBar~\cite{Lees:2014xha}, KLOE~\cite{Archilli:2011zc,Babusci:2012cr,Babusci:2014sta,Anastasi:2016ktq}, and NA48/2~\cite{Batley:2015lha}) and old  beam dump:  E774~\cite{Bross:1989mp}, E141~\cite{Riordan:1987aw}, E137~\cite{Bjorken:1988as,Batell:2014mga,Marsicano:2018krp}), $\nu$-Cal~\cite{Blumlein:2011mv,Blumlein:2013cua},
and CHARM~\cite{Gninenko:2012eq}.
Bounds from supernovae~\cite{Chang:2016ntp} and $(g-2)_e$~\cite{Pospelov:2008zw} are also included.

Colored curves are projections for existing and proposed experiments  on the massive dark photon for $m_{A^{\prime}} >1$ MeV:
Belle-II~\cite{Kou:2018nap} at SuperKEKb;
LHCb upgrade~\cite{Ilten:2016tkc,Ilten:2015hya} at the LHC; 
NA62 in dump mode~\cite{NA62:dump} and NA64(e)$^{++}$~\cite{NA64:eplus} at the SPS; FASER and FASER2~\cite{Feng:2017uoz} at the LHC;
DarkQuest~\cite{Berlin:2018pwi} at Fermilab; HPS~\cite{Adrian:2018scb} at JLAB;
DarkMESA~\cite{Doria:2019sux} at Mainz, and Mu3e experiment at PSI~\cite{Echenard:2014lma}.
For masses above $\sim$100 GeV, projections are obtained for ATLAS/CMS  during the high luminosity phase of the LHC~\cite{Curtin:2014cca}.

\vskip 2mm
From \textbf{Figure~\ref{fig:DP_visible}} it is apparent that collider-based and beam-dump experiments cover a fully complementary region in the parameter space, being collider experiments mostly sensitive to relatively large couplings and masses, and beam-dump experiments sensitive 
to much lower couplings and masses below few GeV.
The motivated range for $\varepsilon$ of $10^{-5} - 10^{-3}$ and masses less than 1~GeV will be covered in the short future mostly by LHCb, HPS, and Mu3e. Proton beam dump experiments (NA62-dump, DarkQuest and possibly SHiP) will push the exploration below $\varepsilon \sim 10^{-5}$.

\subsubsection{Millicharged Particle Production}
\label{ssec:vector_exp_milliQ}
Milli-charged particles arise, as discussed in  Section~\ref{sssec:vector_portal}, in the case of a massless dark photon because the rotation of the mixing term leaves the photon coupled to the dark sector particles $\chi$  with strength $\varepsilon e^\prime$. Searches are accordingly parameterized in terms of the mass $m_\chi$  and the electromagnetic coupling (modulated by  $\varepsilon$) of the supposedly milli-charged dark-sector particle.

The physics of stellar evolution for
horizontal branches, red giants, and white dwarves (RGWD~\cite{Vogel:2013raa}), 
together with supernovae (SN1987~\cite{Chang:2018rso}) 
provide bounds in the region of small masses ($m_\chi < 1$ MeV).
In this region constraints on $N_{eff}$ during nucleosynthesis  and in the cosmic microwave background ($N_{eff}$ BBN and CMB~\cite{Vogel:2013raa}) 
limits the possibility of having  milli-charged particles.

\begin{figure}[h]
\includegraphics[width=12cm]{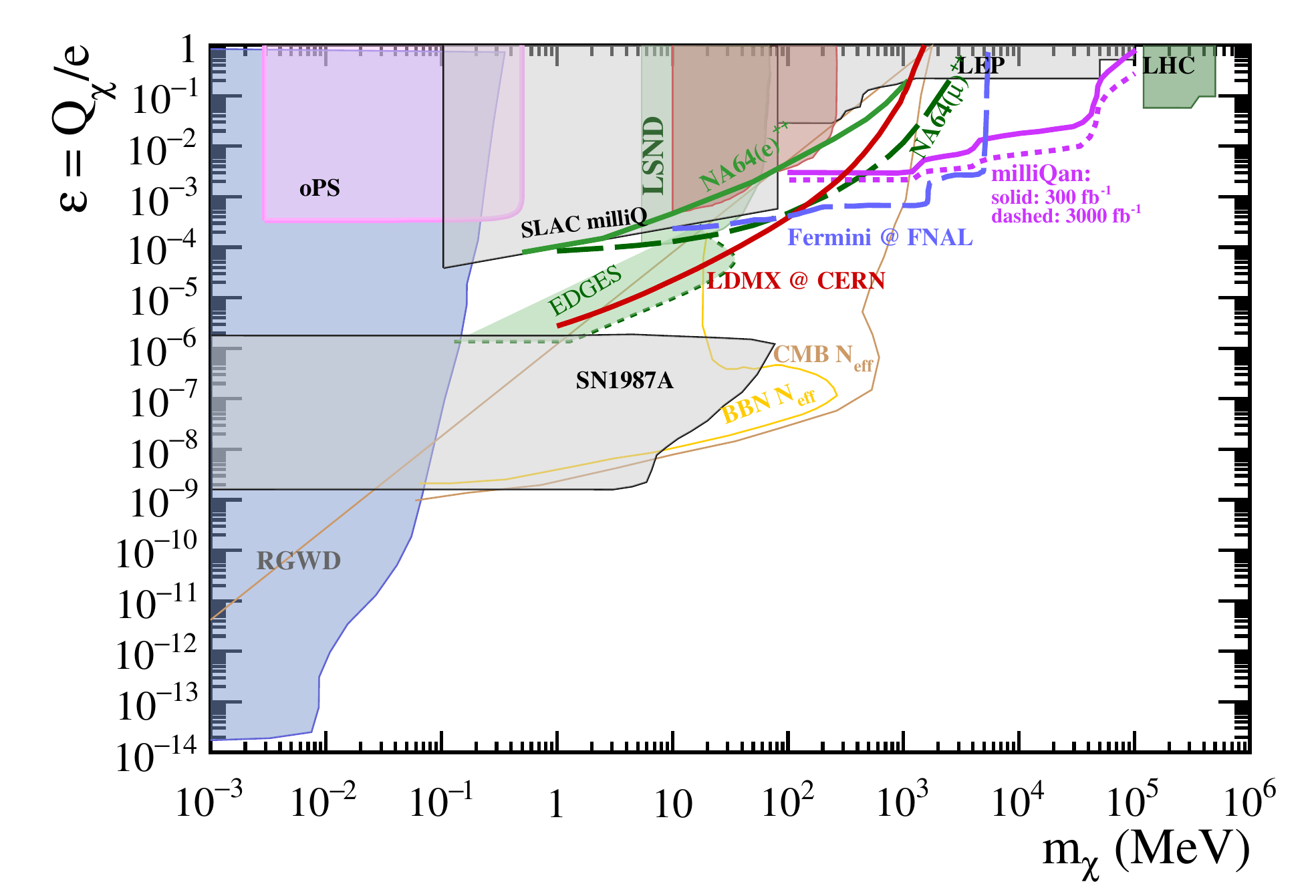}
\caption{\small  \label{fig:milli} {\bf Milli-charged particles.}  
Existing limits (filled areas) and future sensitivities for existing or proposed experiments (curves).
Existing limits: stellar evolution (RGWD~\cite{Vogel:2013raa} and SN1987~\cite{Chang:2018rso});
$N_{eff}$ during BBN and CMB~\cite{Vogel:2013raa};
invisible decays of ortho-positronium (oPS)~\cite{Badertscher:2006fm};
SLAC milliQ experiment~\cite{Prinz:1998ua};
reinterpretation of data from LSND and MiniBooNE~\cite{Magill:2018tbb};
interpretation of the anomalous 21~cm hydrogen absorption signal by EDGES~\cite{Kovetz:2018zan};
searches at LEP~\cite{Davidson:2000hf} and LHC~\cite{Jaeckel:2012yz}.
Future sensitivities:
 NA64(e)$^{++}$~\cite{NA64:eplus}; NA64($\mu$)~\cite{NA64:2018iqr}; FerMINI~\cite{Kelly:2018brz};
 milliQAN~\cite{Ball:2016zrp}; LDMX~\cite{Akesson:2018vlm}. 
 See text for details.  Figure revised from Ref.~\cite{Fabbrichesi:2020wbt}.}
\label{fig:DP_milliQ}
\end{figure}

Further limits can be derived  from precision measurements in QED, notably from the Lamb shift in the transition $2S_{1/2}$-$2P_{3/2}$ in the Hydrogen atom~\cite{Hagley:1994zz}
and the non-observation of the invisible decay of ortho-positronium (oPS~\cite{Badertscher:2006fm}).
Limits in the intermediate mass range $1-100$~MeV come from a SLAC dedicated experiment (SLAC milliQ~\cite{Prinz:1998ua}) and from the reinterpretation of data from the neutrino
experiments LSND and miniBooNE~\cite{Magill:2018tbb}.
Searches at LEP~\cite{Davidson:2000hf} and LHC~\cite{Jaeckel:2012yz} cover larger values of the mass (100 MeV $< m_\chi < 1$~TeV). 
All these limits are shown as filled area in the plot of \textbf{Figure~\ref{fig:milli}}. Milli-charged particles as dark matter have been proposed (see for example~\cite{Kovetz:2018zan} and~\cite{Liu:2019knx}) to explain the anomalous 21~cm hydrogen absorption signal reported by the EDGES experiment~\cite{Monsalve:2018fno}.

The projected limits of future experiments are depicted in \textbf{Figure~\ref{fig:milli}} as coloured curves. 
Of these, the most significative for masses around 1~GeV comes from the proposed milliQAN experiment~\cite{Ball:2016zrp} proposed to be installed on the surface above one of the LHC
interaction points. MilliQAN could improve the collider limits by two orders of magnitude.
The range in mass between 10-100~MeV can be optimally covered by the FerMINI experiment~\cite{Kelly:2018brz}
proposed in the DUNE near detector hall at Fermilab. Finally the search for milli-charged particles 
below 10~MeV mass may be improved by almost two orders of magnitude by the LDMX experiment~\cite{Akesson:2018vlm}
proposed both at CERN~\cite{LDMX:CERN} and at SLAC~\cite{Raubenheimer:2018mwt}.

\clearpage
\subsection{Scalar Portal: search for light scalar particles mixing with the Higgs}
\label{ssec:scalar_exp}

\vskip 2mm
In a minimal scalar portal model, the new singlet scalar has a quadratic coupling $\lambda_{\rm HS}$ and a mixing term to the Higgs $\sin \theta$, as detailed in Section~\ref{sssec:scalar_portal}.
The light scalar production always occurs via the Higgs boson, both on- and off-shell. If the Higgs is on-shell, the production is dominated by the quartic coupling, $\lambda_{\rm HS}$,  via the direct decay $H \to SS$. This process is instead sub-dominant when the Higgs is off-shell. In this case the production occurs via Higgs penguin diagrams, as for example $K \to \pi S$ and $B \to K S$ decays, with a probability proportional to $ \sin^2 \theta$. The mixing term drives also the light scalar decays. In fact, the light dark scalar couples to SM particles as a SM Higgs, but with strength reduced by a factor of $\sin \theta$.

The production processes define also the maximum light scalar mass that can be explored at different experimental facilities: up to $m_K - m_{\pi}$  for kaon factories and proton beam dump experiments (if the the centre-of-mass energy $\sqrt{s}$ is sufficient for production of a pair of strange particles); up to $m_B - m_K$ for $b-$factories, LHCb, and proton beam dump experiments (if $\sqrt{s}$ is above twice the $b$ quark mass), and up to $m_h/2$ for LHC-based experiments. 

\begin{figure}[h]
\includegraphics[width=12cm]{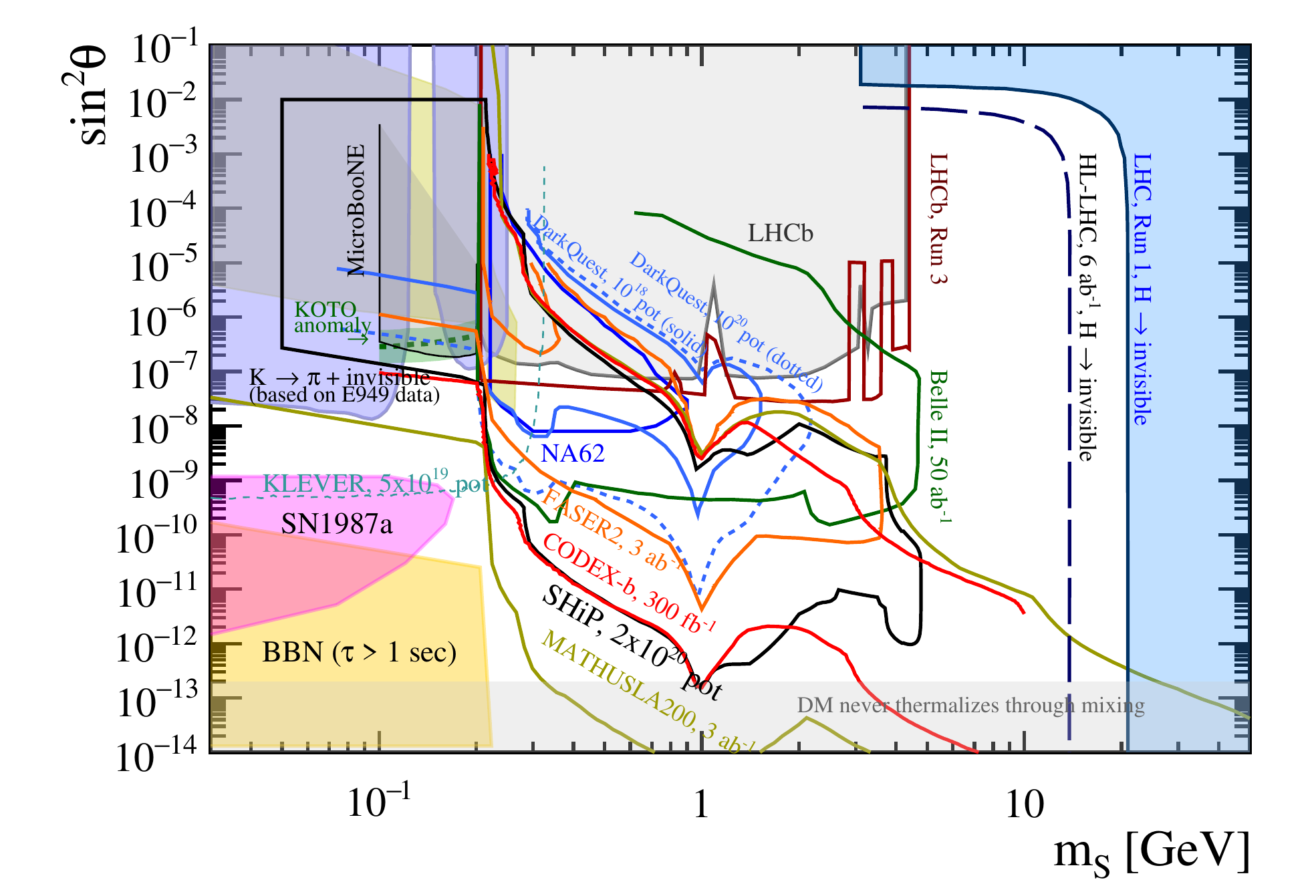}
\caption{ {\bf Dark scalar into visible final states}.
Shaded areas come from: reinterpretation~\cite{Winkler:2018qyg} of results from CHARM experiment~\cite{Bergsma:1985qz}; E949~\cite{Artamonov:2008qb}; MicroBooNE~~\cite{microboone} that excludes a light dark scalar as interpretation~\cite{koto_anomaly_inter} of the KOTO anomaly; LHCb~\cite{Aaij:2016qsm, Aaij:2015tna} and Belle~\cite{Wei:2009zv}.
Coloured lines come from projections of existing/proposed experiments: 
NA62-dump~\cite{NA62:dump} and DarkQuest~\cite{Batell:2020vqn}, Belle II~\cite{Filimonova:2019tuy}, SHiP~\cite{Anelli:2015pba}, FASER2~\cite{Ariga:2018uku}, CODEX-b~\cite{Aielli:2019ivi}, MATHUSLA~\cite{Alpigiani:2020tva}, and KLEVER~\cite{Beacham:2019nyx}. Vertical lines come from the knowledge for the invisible Higgs width after Run~1 at the LHC and projections in the HL-LHC era (see~\cite{Strategy:2019vxc} and references therein). BBN and SN 1987A are from ~\cite{Fradette:2017sdd} and ~\cite{Krnjaic:2015mbs}, respectively.}
\label{fig:DS_1}
\end{figure}

\vskip 2mm
The current experimental bounds and future projections in the plane $\sin^2 \theta $ versus $m_S$  are shown in \textbf{Figure~\ref{fig:DS_1}}.
First investigation of light dark scalar with mass below the kaon threshold has been obtained by reinterpreting~\cite{Winkler:2018qyg} the results from the CHARM experiment~\cite{Bergsma:1985qz}.
Masses below the kaon mass can be optimally explored at kaon factories: E949~\cite{Artamonov:2008qb}, NA62\footnote{See talk from J. Swallow at the FIPs 2020 workshop, https://indico.cern.ch/event/864648/.} and possibly in the future by the KLEVER~\cite{Beacham:2019nyx}) experiment, as by-product of the $K \to \pi \nu \overline{\nu}$ analyses using the missing mass technique and assuming the 2-body decay $K \to \pi S$, where $S$ is not detected. NA62 and KLEVER  should be able to close the gap between the current E949 bound and constraints arising from the supernova SN 1987A (see discussion below).
MicroBooNE~\cite{microboone} recently set a bound between 100 and 200~MeV that excludes a light feebly-interacting scalar as responsible of the KOTO anomaly\footnote{At the KAON 2019 conference, https://indico.cern.ch/event/769729/, the KOTO collaboration announced a preliminary observation of 4 events with background expectation of $0.05\pm 0.02$ in the search for $K_L \to \pi^0 \nu \overline{\nu}$ decay.}~\cite{koto_anomaly_inter}.

\vskip 2mm
Light scalars with masses between the kaon and the $B$ meson mass have been searched for at LHCb~\cite{Aaij:2016qsm, Aaij:2015tna} and Belle~\cite{Wei:2009zv} via the decays $B^{+,0} \to K^{+,*0} S$, $S \to \mu^+ \mu^-$. Both LHCb and Belle II are expected to improve over these limits in the coming years.
In the same mass range, in the close future, NA62-dump~\cite{NA62:dump} and DarkQuest~\cite{Batell:2020vqn}, and, on a longer timescale, Belle~II~\cite{Filimonova:2019tuy} with 50~ab$^{-1}$ and possibly SHiP~\cite{Anelli:2015pba}, will be able to push further down the exploration of the parameter space towards smaller couplings. Experiments at the LHC, as FASER2~\cite{Ariga:2018uku}, CODEX-b~\cite{Aielli:2019ivi}, MATHUSLA~\cite{Alpigiani:2020tva}, and ANUBIS~\cite{Bauer:2019vqk} will possibly extend the sensitivity
for dark scalars up to half of the Higgs mass.
For LHC-based detectors, projections are obtained assuming the quartic coupling fixed to $\lambda_{\rm HS} \sim 6 \cdot 10^{-4}$, which corresponds to the precision ($\sim 1\%$) obtainable on the invisible Higgs branching fraction by a future $e^+ e^-$ circular collider~\cite{Strategy:2019vxc}.

\vskip 2mm
In Figure~\ref{fig:DS_1} the jump in sensitivity at $\sim 200 $~MeV and then at $\sim 3.8~$~GeV is due to the opening of the $\mu^+ \mu^-$ and $\tau^+ \tau^-$ thresholds, respectively, as this portal inherits from the SM its Yukawa structure. The complicated structure at  $\sim$1~GeV is defined by the inclusions of resonant $\pi \pi$ final states whose rate is still known only with large uncertainties~\cite{Winkler:2018qyg}. 

\vskip 2mm
Finally, in the limit of small mixing angle, one can bound the quartic coupling $\lambda_{\rm HS}$ via the Higgs invisible branching fraction, since $\lambda_{\rm HS}$ is naturally expected to satisfy the relation $\lambda_{\rm HS} \leq m^2_S/v^2$. Vertical lines correspond to the bounds obtainable on $\lambda_{\rm HS}$ from the knowledge for the invisible Higgs branching fraction from ATLAS and CMS after LHC Run~1 and projections in the HL-LHC era (see~\cite{Strategy:2019vxc} and references therein).

\vskip 2mm
As discussed in Section~\ref{ssec:astroparticle}, light dark scalars can be also constrained by astrophysics sources.
A light ($<$ 100~MeV) dark scalar with a thermal number density can decay during BBN and spoil the successful predictions of light element yields accumulated in the early Universe~\cite{Fradette:2017sdd}. Likewise, a light dark scalar produced on shell during a supernova (SN) explosion can significantly contribute to its energy loss, thereby shortening the duration of the observable neutrino pulse emitted during core collapse (see for example, \cite{Krnjaic:2015mbs}).

\vskip 2mm
Light dark scalars can be also mediators between DM $\chi$ and SM particles. However has been demonstrated~\cite{Krnjaic:2015mbs, deNiverville:2012ij} that for mediator masses  $m_S \geq 2 m_{\chi}$,  DM annihilating directly into SM particles is ruled out by low-energy experiments for nearly all DM candidates
under the most conservative assumptions regarding the
DM-mediator couplings and mass ratios. 
If the mediator is lighter than $ m_{\chi}$ and the relic abundance is set by secluded annihilation $ \chi \overline{\chi}\to SS$, the $\sin \theta$ mixing parameter 
is only bounded from below by the DM thermalization requirement ($\sin^2 \theta > 10^{-13}$)~\cite{Krnjaic:2015mbs} and the parameter space is still open for exploration.

\vskip 2mm
Non-minimal models that could explain the Higgs hierarchy problem (Neutral Naturalness, Twin Higgs, relaxion) or originate a strong EW phase transition, 
include the scalar portal as connection between NP and SM particles. These models have similar phenomenology as the minimal one and can provide important target areas for future searches.

\subsection{Pseudo-Scalar Portal: search for heavy axions and ALPs}
\label{ssec:pseudo_scalar_exp}

The pseudo-scalar portal (see Section~\ref{sssec:pseudo_scalar_portal}) embraces any type of pseudo-Nambu-Goldstone bosons (pNGB) of a spontaneously broken $U(1)$ symmetry at a scale $f_a$.
The principal example of very light pNGB is the axion 
but natural extensions of the axion paradigm bring to a wide range of interesting pseudoscalar particles which typically have very similar interactions as the axion, but without a strict relation between the mass and the coupling, the Axion-Like Particles or ALPs.
ALPs also provide an interesting connection to the puzzle of dark matter, because they can mediate the interactions between the DM particle and SM states and allow for additional annihilation channels relevant for the thermal freeze-out of DM. In fact in presence of an additional pseudoscalar particle mediating the interactions of DM with the SM sector, 
constraints from direct detection experiments 
and invisible Higgs width that affect the scalar portal can be easily evaded. 
Moreover, if the pseudo-scalar mass is less than twice the mass of the DM particle $\chi$, the annihilation process $\chi \chi \to a a$, followed by decays of the pseudo-scalars into SM particles, allows for a highly efficient annihilation of DM particles. The only important constraint is that such pseudo-scalar particles must decay before BBN. 

\vskip 2mm
As explained in Section~\ref{sssec:pseudo_scalar_portal}, ALPs can mediate interactions between DM and the SM sector via three different couplings, photon-, gluon-, and fermion-coupling.
In this study only experimental results related to photon-coupling are considered.
In this case all the ALP phenomenology
(production, decay, oscillation in magnetic field) is fully defined in the ($m_a$; $|g_{a\gamma}|  = f^{-1}_{a}$) parameter space.

\begin{figure}[h]
\includegraphics[width=12cm]{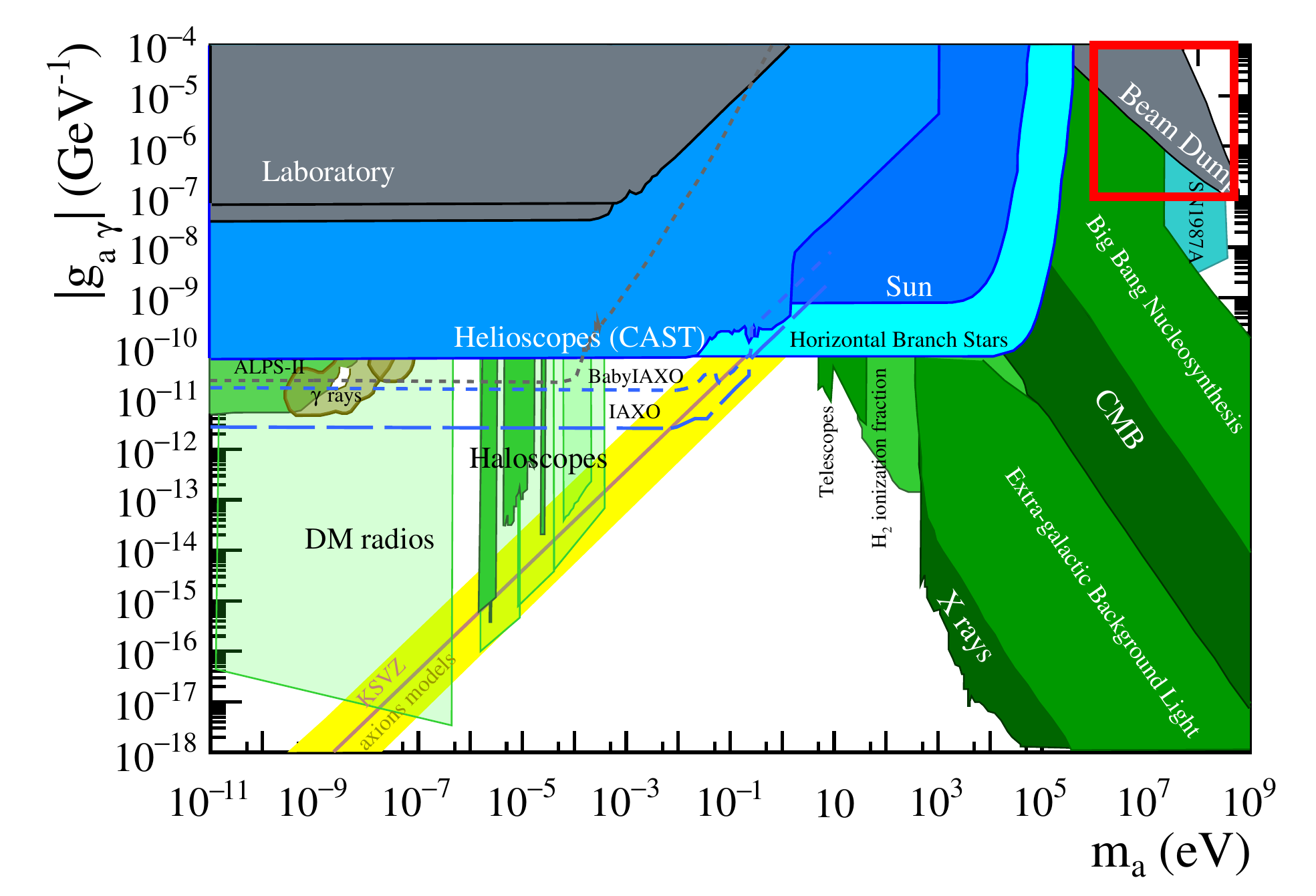}
\includegraphics[width=12cm]{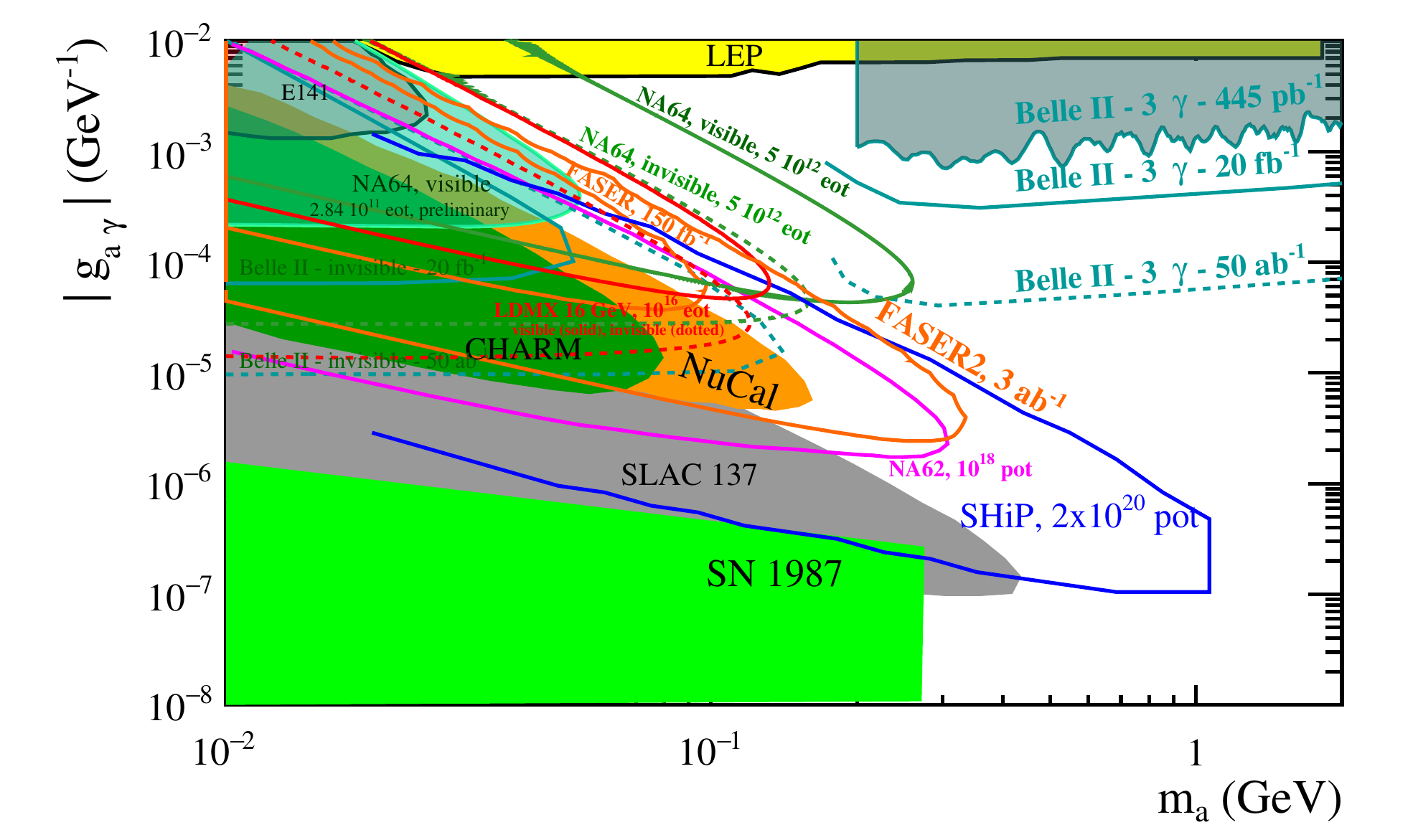}
\caption{{\bf Axions/ALPs with photon coupling.} {\it Top plot:} overall view of axion/ALP searches. Figure revised by I.~Irastorza from Ref.~\cite{Beacham:2019nyx}. {\it Bottom plot:} zoom in the region of interest for accelerator-based experiments (note the different units for the mass ranges in the two plots). 
Shaded areas are excluded regions from:
LEP (data:\cite{Acciarri:1994gb,Abreu:1991rm, Abreu:1994du, Acciarri:1995gy}; interpretation:~\cite{Jaeckel:2015jla};
Belle II~\cite{BelleII:2020fag};
E141 (data~\cite{Riordan:1987aw} and interpretation~\cite{Dobrich:2017gcm}); 
E137~\cite{Bjorken:1988as}; 
NA64~\cite{Banerjee:2020fue};
CHARM~\cite{Gninenko:2012eq}; 
NuCal~\cite{Blumlein:1990ay}
Curves are projections from:
Belle II~\cite{Kou:2018nap} for 20~fb$^{-1}$ and 50~ab$^{-1}$;
SHiP~\cite{Anelli:2015pba};
FASER and FASER2~\cite{Beacham:2019nyx};
NA64$_e^{++}$~\cite{NA64:eplus} in visible and invisible modes.
}
\label{fig:PS_photon}
\end{figure}

\vskip 2mm
Searches for axions with photon couplings in the sub-eV mass range have literally exploded in the recent past. 
The most updated review on the laboratory searches for very light axions and ALPs is contained in Ref.~\cite{Irastorza:2018dyq} and then updated in~\cite{Beacham:2019nyx}.
\textbf{Figure~\ref{fig:PS_photon}} 
shows the sensitivity of current and future experiments to axions/ALPs with photon coupling over about 20 orders of magnitude in mass and $\sim$~14 orders of magnitude in coupling.
The Figure follows a colour scheme to present results obtained with different methods: black/grey for laboratory results, bluish colours for searches for axions from the sun (helioscopes) and bounds related
to stellar physics, greenish for searches for axions as potential DM candidates (haloscopes) or cosmology dependent arguments (see \cite{Beacham:2019nyx} for details). Hinted regions, like the QCD axion, are in yellow/orange.
Laboratory limits (dark grey areas) are essentially due to the results of OSQAR~\cite{Ballou:2015cka}  (region below 1 meV), and PVLAS~\cite{DellaValle:2015xxa} (region above 1 meV).
The bounds from helioscopes and haloscopes experiments are mostly driven by CAST~\cite{Anastassopoulos:2017ftl} and ADMX~\cite{Du:2018uak} results.
Projections for laboratory experiments, helioscopes, and haloscopes,  are dominated by ALPS-II~\cite{Bahre:2013ywa}, IAXO~\cite{Armengaud:2014gea} (and its smaller version BabyIAXO), DM-radio~\cite{Silva-Feaver:2016qhh} and ADMX and MADMAX~\cite{Brun:2019lyf}, and they are shown as gray, blue, and green dashed curves, respectively.

\vskip 2mm
The contribution from accelerator-based experiments below a few GeV mass scale is highlighted in the red box in Figure~\ref{fig:PS_photon} (top)  and then zoomed in Figure~\ref{fig:PS_photon} (bottom). This region of parameter space is particularly interesting because it covers the motivated MeV-GeV region still compatible with the QCD axion band. 
Searches for ALPs with photon coupling have been performed at $e^+e^-$ collider-based experiments  using mono-photon  searches ($e^+ e^- \to \gamma^* \to a \gamma$) at LEP (data~\cite{Acciarri:1994gb,Abreu:1991rm, Abreu:1994du, Acciarri:1995gy}; interpretation~~\cite{Jaeckel:2015jla}) and, recently, tri-photon searches in the process $e^+ e^- \to \gamma a, a \to \gamma \gamma$ at Belle II~\cite{BelleII:2020fag} with $\sim 0.5$ fb$^{-1}$ of integrated luminosity.

In electron beam dump experiments (E141: data~\cite{Riordan:1987aw} and interpretation~\cite{Dobrich:2017gcm}; E137~\cite{Bjorken:1988as}; NA64~\cite{Banerjee:2020fue} ) and proton beam dump experiments (CHARM~\cite{Gninenko:2012eq}; NuCal~\cite{Blumlein:1990ay}), the ALPs are supposed to be produced mainly via the Primakoff effect, i.e. the conversion of a photon into an ALP in the vicinity of a nucleus~\cite{Halprin:1966zz}.
In the close future, NA62 in dump mode, NA64$_e^{++}$~\cite{NA64:eplus}, and FASER should be able to improve over the current results for masses below 1~GeV.

ATLAS and CMS are currently searching for ALPs with photon couplings in a relatively higher mass region, $5 <m_a < 100s$~GeV, using  exclusive di-photon final states (as for example in the  process $pp,PbPb \to  \gamma \gamma  \to a (\gamma \gamma)$~\cite{Knapen:2016moh, Sirunyan:2018fhl, Aad:2020cje};
inclusive $\gamma \gamma$ resonances (as in $pp \to Z \to  a (\gamma \gamma)) \gamma$)~\cite{Aaboud:2016tru}, 
and exotic $Z$ or Higgs decays (as $pp \to H \to a(\gamma \gamma)+Z,\gamma$ or $pp \to H \to a(\gamma \gamma) + a(\gamma \gamma$)~\cite{Aad:2015bua, Bauer:2017ris, Knapen:2017xzo}.

Constraints on the photon-ALP coupling arise also from the study of the neutrino signal from the supernovae SN1987A where axions or ALPs with masses up to about 100~MeV, possibly copiously produced in the hot core of a supernova, could constitute a new energy loss mechanism~\cite{Dolan:2017osp}.


\subsection{Fermion Portal: search for Heavy Neutral Leptons}
\label{ssec:fermion_exp}
Heavy Neutral leptons (HNLs) can be produced in any process involving neutrinos in the final states, such as pion, kaon, $D$, $B$~mesons, and $Z$ decays, with a probability proportional to the mixing parameters $|U_{e,\mu,\tau}|^2$. Once produced, the HNLs can decay via Charged Current  and Neutral Current  interactions into active neutrinos and other visible final states, as pions, muons and electrons.

Heavy neutrinos mixing with the first and second lepton generation, $\nu_e$ and $\nu_{\mu}$, can be searched for in the missing mass distribution of pions and kaons leptonic decays, eg.$ K,\pi \to \mu (e) ^+ \nu_{\mu} (\nu_{e})$. These searches are very robust because assume only that a HNL mixes with $\nu_e$  and $\nu_{\mu}$, without any further assumption. 
Another strategy to search for HNLs 
is via searches of their decay products. This is the typical technique used in proton beam-dump experiments where the HNL production process is not known. 
Results obtained with this technique are in general less robust than those using the missing mass technique, as they would be largely weakened if the HNLs have unknown decay modes into invisible particles.
Other ways to constrain the couplings of HNLs in a relatively high mass regime is using possible $Z^0$ decays into heavy neutrinos from LEP data. 
In this case only large values of the mixing angles can be explored.

\vskip 2mm
The above techniques assume that HNLs are on-shell particles that can be seen either via visible decays or using the missing mass technique. These are commonly called {\it direct searches}.   However HNLs can be also searched for via precision tests or searches for rare processes in the SM that are indirectly affected through the modification of the active neutrino interactions via the mixing with HNLs ({\it indirect searches}). In fact, due to the presence of sterile neutrinos, the mixing matrix of the three active neutrinos is effectively non-unitary. The subsequent modification of the weak currents then leads to modified predictions for electroweak observables (as, for example, the Fermi constant $G_F$ or the Weinberg angle $\sin \theta_W$) compared to the SM. However, in the mass range considered in this work, 0.1-100~GeV, direct searches dominate the sensitivity.

\begin{figure}[h]
\includegraphics[width=12cm]{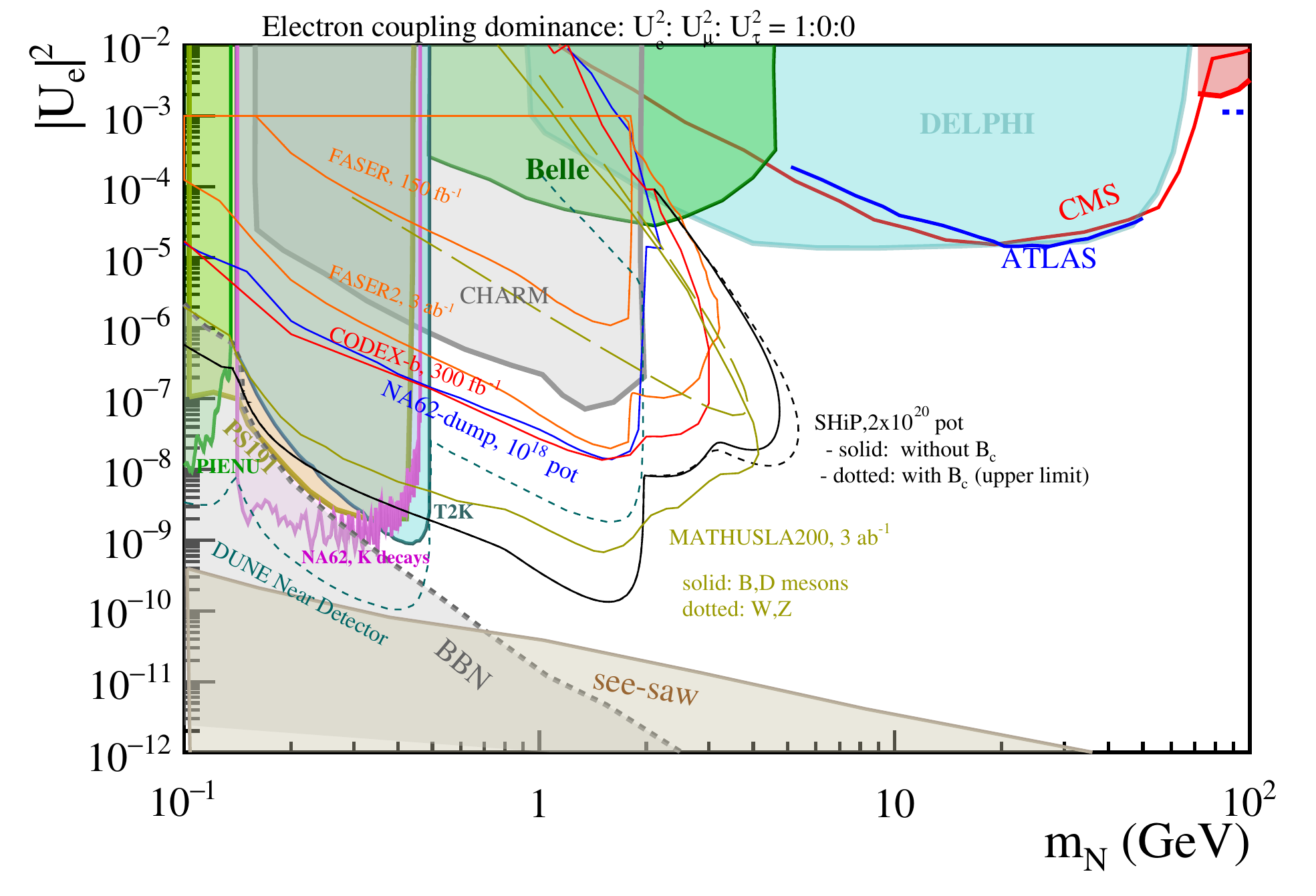}
\includegraphics[width=12cm]{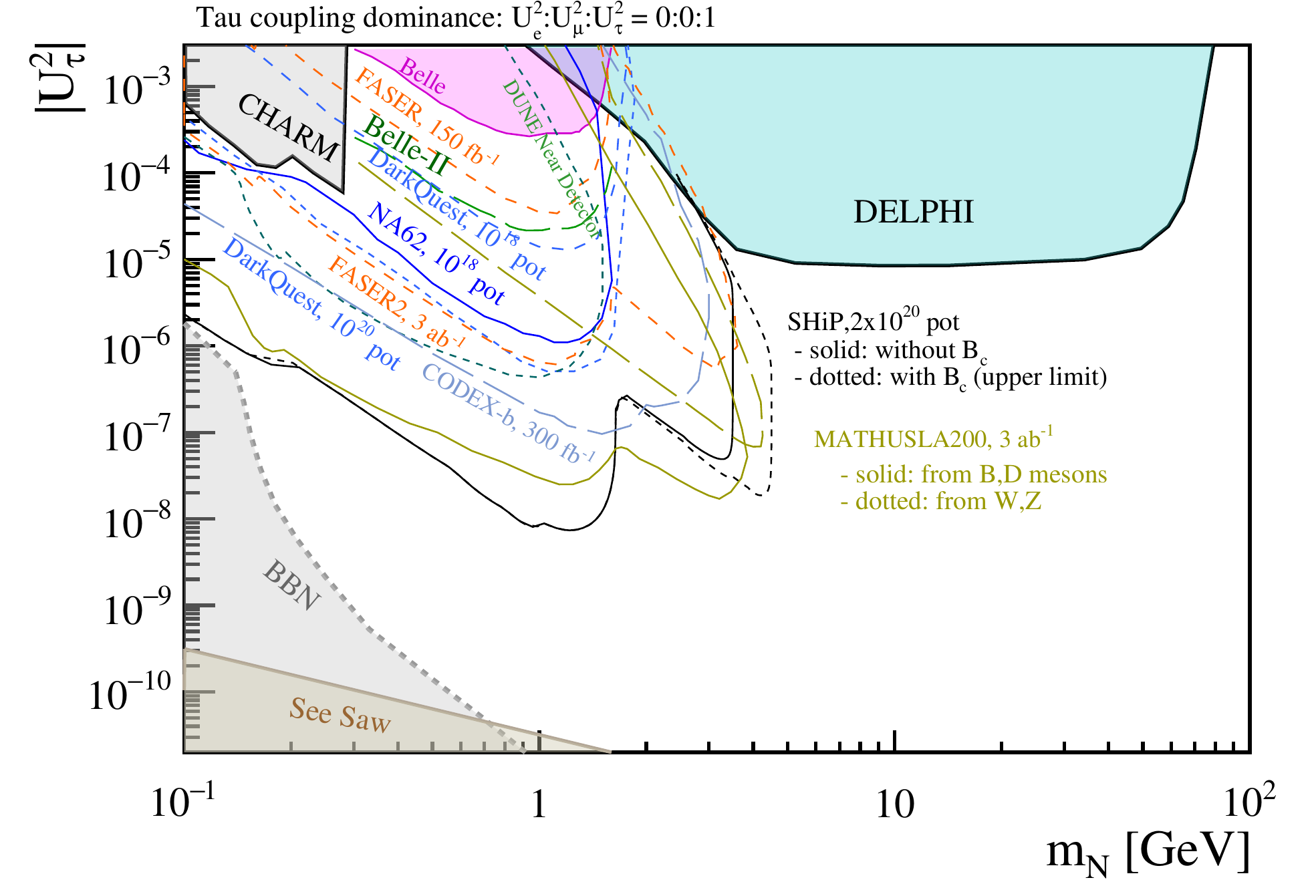}
\caption{{\bf Heavy Neutral Leptons} with coupling to the first (top) and third (bottom) lepton generation. Filled areas are existing bounds from:
PS191~\cite{Bernardi:1987ek}, CHARM~\cite{Bergsma:1985qz}, 
PIENU~\cite{Aguilar-Arevalo:2017vlf}, Belle~\cite{Liventsev:2013zz}; DELPHI~\cite{Abreu:1996pa}, ATLAS~\cite{Aad:2019kiz} and CMS~\cite{Sirunyan:2018mtv}.
Coloured curves are projections from:  NA62-dump~\cite{NA62:dump, Beacham:2019nyx},
DarkQuest~\cite{Batell:2020vqn},
Belle-II~\cite{Dib:2019tuj},
FASER and FASER2~\cite{Feng:2017uoz}; SHiP~\cite{Anelli:2015pba}, DUNE near detector~\cite{Ballett:2019bgd}, CODEX-b~\cite{Aielli:2019ivi}, and MATHUSLA200~\cite{Alpigiani:2020tva}.
BBN and seesaw bounds are computed under the hypothesis of two HNLs mixing with active neutrinos, and should be considered only indicative (see text).}
\label{fig:HNL}
\end{figure}

\vskip 2mm
In the current study we assume, for simplicity,  that only one coupling is switched on at the time. Future studies will include combinations of coupling ratios
compatible with the active neutrino mixing parameters (see, eg, ~\cite{Drewes:2018gkc, Caputo:2017pit}).
\textbf{Figure~\ref{fig:HNL}}  show the current bounds and sensitivity projections for HNLs coupled to the first (top) and third (bottom) lepton flavor generation, respectively.

\vskip 2mm
Strong constraints on couplings for HNLs with masses below the kaon mass are set by past experiments, in particular PS191~\cite{Bernardi:1987ek}, CHARM~\cite{Bergsma:1985qz}, NuTeV~\cite{Vaitaitis:1999wq}, E949~\cite{Artamonov:2014urb},
PIENU~\cite{PIENU:2011aa},  TRIUMF~\cite{Britton:1992xv}, and NA3~\cite{Badier:1985wg}. 
A new analysis of $10^7 \pi^+ \to e^+ \nu_e$ decays by the PIENU collaboration~\cite{Aguilar-Arevalo:2017vlf} has produced the most stringent limit of $|U|^2_e$ below the pion threshold.
The off-axis near detector of the T2K experiment ND280~\cite{Abe:2019kgx} with 
($12.34 \times 10^{20}$~pot ($\nu$-mode) and $6.3  \times 10^{12}$ pot (anti-$\nu$ mode)), and the NA62 experiment with $3.5 \times 10^{12}$  $K^+ \to e^+ \nu_e$ decays in the fiducial volume~\cite{NA62:2020mcv}
have pushed the exploration of the HNL couplings down to $\sim 10^{-9}$ level for masses up to the Kaon mass.

It is important to note that both the BBN and seesaw bounds are meaningful only if more than one HNL-active neutrino coupling is at work, and should not be considered for the single-flavor dominance case studied in this manuscript. However they are included in Figures~\ref{fig:HNL} to show that all the couplings are bounded from below as soon as two HNLs are at work.

Above the kaon mass the current constraints on $|U^2_e|$ weaken significantly,
being dominated by results from the old CHARM experiment (up to the charm threshold), by Belle~\cite{Liventsev:2013zz} up to the $b$ threshold and by DELPHI~\cite{Abreu:1996pa}, ATLAS~\cite{Aad:2019kiz} and CMS~\cite{Sirunyan:2018mtv} up to the $Z$ threshold.

\vskip 2mm
In five year timescale, NA62-dump~\cite{Drewes:2018gkc} and DarkQuest~\cite{Batell:2020vqn} (only for $|U^2_{\tau}|$) can improve the bounds from the CHARM experiment below the charm threshold and be competitive with the projections for FASER and FASER2~\cite{Feng:2017uoz}, while  a significant improvement in the entire mass range below the $B-$meson mass could be achieved on a longer time scale by SHiP~\cite{Anelli:2015pba}, Belle-II~\cite{Dib:2019tuj}, DUNE near detector~\cite{Ballett:2019bgd}, CODEX-b~\cite{Aielli:2019ivi}, and MATHUSLA200~\cite{Alpigiani:2020tva}.  On a much longer timescale experiments at future high-energy hadron and lepton colliders will be sensitive to HNLs well beyond the $b-$meson mass, via direct and indirect searches, see~\cite{Strategy:2019vxc} for a summary.

\clearpage
\section{CONCLUSIONS AND OUTLOOK}
\label{sec:conclusions}
In recent years, the physics of feebly-interacting particles and dark sectors has received considerable and growing attention from the broad HEP community, motivated by both existing data, and by the appeal of simple extensions of the Standard Model to address long-standing puzzles in which FIPs play an important role. While a broad mass range is interesting, special focus has been placed on the relatively poorly explored MeV-GeV scale, for several reasons, and a systematic investigation of this mass range is now under way. To-date, the research programmes in this direction have largely occurred in a manner parasitic to other research goals, or have been focused on the reinterpretation of old data sets. Relatively few dedicated experiments are currently running, though the number planned and expected to run in the coming years has grown considerably. Nevertheless, the wealth of results that have been produced in the past 5-10 years already allow us to draw several noteworthy conclusions:

\begin{itemize}
\item Results from the dedicated experiment NA64$_e$ as well as recent proton beam dump searches for light DM in the MeV-GeV range have begun to exclude new parameter space below 100 MeV in certain specific (but well motivated) models where direct thermal freeze-out can explain the DM abundance (Figure~\ref{fig:DP_y_scalar}); 

\item Dedicated searches for dark photons have excluded the interesting parameter range where a dark photon could be responsible for the $(g-2)_{\mu}$ anomaly (Figure~\ref{fig:DP_visible}). 

\item  HNLs coupled to the first and second SM lepton generation and with masses below the Kaon mass are excluded by NA62 and T2K searches down to $10^{-9}$ (Figure~\ref{fig:HNL}). These bounds, computed under the simplistic assumption of single-flavor dominance, are excluding most of the parameter space below the kaon mass allowed by BBN and seesaw constraints.

\item Heavy axion-like particles with masses between 10 and few hundreds MeV are excluded for photon couplings compatible with the QCD axion band (Figure~\ref{fig:PS_photon}).

\item A light dark scalar acting as the mediator between the SM and DM in $s$-channel thermal freeze-out DM annihilation has been ruled out - in its minimal form - by low-energy experiments running at $b-$ or Kaon factories. Significant parameter space for secluded annihilation  ($ \chi \chi\to $pair~of~mediators$\to$SM) is still open above the cosmological bound of $\sin^2 \theta > 10^{-13}$ (Figure~\ref{fig:DS_1}).

\end{itemize}

In the next ten years a multitude of new initiatives (or upgrades of ongoing experiments) will produce a large set of results applicable to FIP and dark sector physics. In parallel, ongoing theoretical work is providing guidance with respect to regions of parameter space in minimal models that are deserving of special attention. Advanced calculations of FIP production and decay rates, and the inclusion of more complete bounds from astrophysical and cosmology data, will all be important for properly interpreting the wave of new data expected in the coming years. Though the future is not always straightforward to predict, we expect several interesting milestones to be reached by upcoming experiments over the next ten years:

\begin{itemize}
\item Most of the parameter space below ${\mathcal O(300)}$~MeV where direct thermal freeze-out can explain the DM abundance (for both scalar and fermion DM models interacting through a light DP) will be explored by dedicated searches, such as LDMX at SLAC, NA64 at CERN, and beam dump experiments at FNAL and JLab. Likewise, Belle-II is expected to probe the region above ${\mathcal O(300)}$~MeV. Additionally, direct detection sensitivity will be extended by a series of new experiments including SENSEI, DAMIC, and Super-CDMS. Taken together, these experiments will broadly extend sensitivity to WIMP-like dark matter deep into the sub-GeV range, analogous to the way that the LHC and large-scale direct detection experiments will probe the 100s GeV-TeV mass range. 
\item Important regions of the parameter space for the QCD axion, and axion-like particles, will be explored in the coming years (Figure~\ref{fig:PS_photon}). In particular, helioscopes will improve sensitivity to axions and ALPs to probe parts of parameter space consistent with recent hints on anomalous energy loss channels in stars and/or on anomalous $\gamma$-ray transparency of the Universe. Haloscopes  will complement this quest by increasing sensitivity down to the range of coupling and mass where the QCD axion can be a dark-matter candidate. Additionally, much of the ALP parameter space where simple misalignment predicts the correct DM density will be explored, for example by ADMX and DM Radio.
Finally, accelerator-based experiments will continue to explore the well-motivated QCD scale (MeV-GeV) and beyond at larger ALP couplings. 

\item searches for HNLs will be performed by NA62-dump, DarkQuest, Belle II and LHC-based experiments, in the still mostly uncharted region above the kaon mass, in a range of couplings compatible with successful leptogenesis. 

\end{itemize}

The field of FIPs and dark sectors is still fresh and many possible avenues can and should be pursued in the future. Theoretical studies should continue to map out how the known portal operators are embedded in complete models addressing the varied puzzles of the SM, and this will be important for identifying additional promising sensitivity milestones around which to focus experimental efforts. At the same time, such studies will naturally motivate the inclusion of less minimal models with more complicated signatures in future searches. Within the context of even the minimal models described in this review, much work remains regarding the comparison of results among experiments. For example, understanding the interplay between active neutrino mixing parameters derived from ongoing and future neutrino experiments and the favoured ranges of HNL parameters would be important, as would more comprehensive comparisons of sub-GeV direct detection searches and accelerator based searches. Likewise, the wealth of data coming from accelerator based experiments should be compared in greater detail to existing and future astrophysics and cosmology driven sources of insight. 

\vskip 2mm
Given the lessons of the LHC, addressing the fundamental and interconnected open questions in particle physics requires today a diversified and cross-frontier research programme incorporating accelerator physics, underground detectors, cosmology and astrophysics instruments, and precision experiments, all supported by a strong and focused theoretical involvement. In this vein, the field of FIPs and dark sector physics is casting a light on the sub-GeV frontier, and offers many opportunities for exciting and profound discoveries in the future. 

\section*{ACKNOWLEDGMENTS}
We acknowledge the use of the {\sc Darkcast} package\footnote{Website: \url{https://gitlab.com/philten/darkcast}}~\cite{Ilten:2018crw} for some of the numerical results shown in this review. GL warmly thanks the BSM Working Group of the Physics Beyond Colliders initiative and the speakers and organizing committee of the FIPs 2020 workshop for the lively discussions about most of the topics covered in this review. MP is supported in part by U.S. Department of Energy (Grant No. desc0011842). PS is supported by the U.S. Department of Energy under Contract No. DE-AC02-76SF00515.


\end{document}